\documentclass[prd,tightenlines,nofootinbib,superscriptaddress]{revtex4}

\usepackage[T1]{fontenc}
\usepackage[utf8]{inputenc}

\usepackage{amsfonts,amssymb,amsthm,bbm,amsmath,wasysym}

\usepackage{hyperref}

\usepackage{subcaption}
\usepackage{color,psfrag}
\usepackage[dvips]{graphicx}

\usepackage{tikz}
\usetikzlibrary{calc}
\usetikzlibrary{decorations.pathmorphing}

\newcommand{\be}{\begin{equation}}
\newcommand{\ee}{\end{equation}}
\newcommand{\beq}{\begin{eqnarray}}
\newcommand{\eeq}{\end{eqnarray}}
\newcommand{\bes}{\begin{eqnarray}}
\newcommand{\ees}{\end{eqnarray}}

\newcommand{\sgn}{\mathrm{sgn}}

\begin{document}

\title{Abelian 2+1D Loop Quantum Gravity Coupled to a Scalar Field}

\author{{\bf Christoph Charles}}\email{c.charles@ipnl.in2p3.fr}
\affiliation{Univ Lyon, Université Lyon 1, CNRS/IN2P3, IPN-Lyon, F-69622, Villeurbanne, France}

\date{\today}

\begin{abstract}
  In order to study 3d loop quantum gravity coupled to matter, we consider a simplified model of abelian quantum gravity, the so-called $\mathrm{U}(1)^3$ model. Abelian gravity coupled to a scalar field shares a lot of commonalities with parameterized field theories. We use this to develop an exact quantization of the model. This is used to discuss solutions to various problems that plague even the 4d theory, namely the definition of an inverse metric and the role of the choice of representation for the holonomy-flux algebra.
\end{abstract}

\maketitle

\tableofcontents

\section{Introduction}

In order to tackle the problem of quantum gravity, instead of studying the full theory of general relativity, it is possible to study simpler models. One such model is pure 3d gravity, which describes a simplified universe with only 2 spatial dimensions and 1 dimension of time and without matter. Since classical 3d gravity is a topological theory (it does not have local degrees of freedom), its quantum theory is much more tractable as was originally noticed by Witten \cite{Witten:1988hc}. Since then, the model has been studied in various other manners, including using Loop Quantum Gravity techniques \cite{Freidel:2002hx,Noui:2004iy}. Several directions can be considered from there. One could use the techniques developed to consider a four-dimensional theory and therefore follow the LQG developments. Or it is possible to try and couple 3d gravity to matter, in order to get a more complete model.

This last direction is however rather difficult since the main property of 3d gravity, namely its topological nature, is generically lost when coupling to matter. In the context of Loop Quantum Gravity, no complete model of 3d gravity coupled to matter, even a simple scalar field, is known \cite{Date:2011bg} \footnote{There is however substantial work trying to use matter as a clock \cite{Giesel:2012rb,Bilski:2017sze}. In that case, the scalar field is used to fix the gauge and the resulting theory is formulated as a diffeomorphism invariant theory. This actually evades the problem of Dirac observable we mention a bit later.}. This is partially due to difficulties in quantizing scalar fields in LQG \cite{Thiemann:1997rq,Ashtekar:2002vh,Kaminski:2005nc,Kaminski:2006ta}, partially due to difficulties in constructing Dirac observables \cite{Dittrich:2004cb} but also simply to the difficulties in writing the Hamiltonian constraints involving an inverse metric \cite{Thiemann:1996ay,Livine:2013wmq}.

It does not mean that no reasonable conjecture is known. A surprising number of elements, at least from an LQG perspective \cite{Freidel:2005bb,Freidel:2005me,Ashtekar:1998ak}, converge towards the idea that spacetime in 3d quantum gravity is best described by a non-commutative manifold when coupled to matter. In this regard, non-commutative field theory (see for instance \cite{Szabo:2001kg}) would be the right effective field theory to describe quantum gravity phenomena, at least in three dimensions. This new non-commutative structure is particularly interesting because it seems to be specific to quantum gravity phenomena and as such, it does provide potential insights for studying the full 4d theory. Our goal in this paper is therefore to work towards the goal of developing a rigorous, non-perturbative theory of 3d quantum gravity coupled to matter (most probably just a scalar field) in the context of LQG. If such a theory can be developed, we will finally be able to test the conjectures regarding the non-commutative structure of spacetime, at least in 3d.

In this paper and as a first step in this project, we will study the quatum theory of matter coupled to 3d \textit{linear} gravity. The \textit{linear} term here refers to the fact that we will consider a simplification on the gravity side, by considering an abelian gauge group (rather than the usual local Lorentz invariance). This model is inspired by Smolin's remark on the $G \rightarrow 0$ limit of gravity (where $G$ is Newton's constant) \cite{Smolin:1992wj}. This model, called the $\mathrm{U}(1)^3$ model, corresponds to the usual linearized gravity theory but expressed in a diffeomorphism invariant manner. This simplification might seem quite drastic, especially in 3d for which linearized gravity is quite trivial. Still, it does serve two purposes. First, pure 3d gravity, which has been studied so far, can be considered a simplification on the matter side. Here, we are trying to keep matter but rather simplify the gravity side in order to get new insights. Second, as we will see, and perhaps unsurprisingly, this linear theory is exactly solvable and exactly quantizable (at least with a few assumptions on the topology). The way it is solved however is interesting. Indeed, by writing every expressions in a diffeomorphism invariant manner, we will get formulas that are starting points for the full theory, either by deforming them accordingly, or as initial point for a perturbative study. On top of these expected benefits, we will also get interesting results and insights on how quantum matter and quantum spacetime interacts. In particular, our work reveals more precisely the role of the BF representation \cite{Dittrich:2014wpa,Bahr:2015bra} of the holonomy-flux algebra with respect to the solutions of the theory but also the role of unconventional representations (inspired from \cite{Koslowski:2007kh,Sahlmann:2010hn,Koslowski:2011vn}) in the construction of the field operators.

The main result of this paper is that, in this simplified setting of a scalar field coupled to 3d linear gravity, two sectors entirely decouple. One of the sector correspond to the matter sector. Its structure is exactly equivalent to the free scalar field though expressed in a diffeomorphism invariant way. The second sector roughly corresponds to gravity and is governed by equations similar to BF theory. This separation is possible because we can write the equivalent of creation and annihilation operators of the free field theory, with the additional property of commuting with all the constraints. The first sector correspond to the states explored by the ladder operators while the second sector correspond to the part on which the constraints act. This separation allows the definition of an explicit exact (though trivial) quantum theory. It is noteworthy however that the scalar field operators (the field operator and its canonically conjugated momentum) cannot be expressed in the natural representations of the algebra we found, even though the ladder operators can. The problem is linked to the definition of the inverse of the determinant of the triad, a problem widely encountered in LQG \cite{Thiemann:1996ay,Livine:2013wmq}. It is possible to solve this problem in this simplified context by appealing to representations that are peaked on classical solutions of the Gauß constraints. This result might indicate a possible route for solving similar problems in non-linear or 4d theories.

The paper is organized as follows. The first section gives a bird eye view on the ideas of the paper, staying quite general but still giving more technical details than this introduction. The second section is devoted to the classical study of the theory, in particular the decoupling of the two sectors classically. The third section is concerned with the quantization of the theory. Two approaches are provided: the naive approach that correspond to the previous study and a second approach that allows the development of all the fundamental operators. Finally, the last section discusses various implications of the results with regard to future work.

\section{Overview}

The model we intend to study in the end is 3d quantum gravity coupled to matter. More specifically here, we want to couple a scalar field to gravity in a quantum theory. For this, we can start from the standard action:
\begin{equation}
S[e,A,\phi] = \int_{\mathcal{S}} \left(\alpha \epsilon_{IJK} e^I \wedge F^{JK}[A] + \frac{\Lambda}{6} \epsilon_{IJK} e^I \wedge e^J \wedge e^K + \frac{1}{2} \star \mathrm{d}\phi \wedge \mathrm{d}\phi + \frac{m^2}{2} \star \phi \wedge \phi \right) .
\end{equation}
Here, $\mathcal{S}$ is the spacetime manifold. $e$ is the triad. It is an $\mathbb{R}^3$-valued $1$-form that can be interpreted as an $\mathfrak{su}(1,1)$-valued one using the Levi-Civita symbol. $A$ is the spin connection. It is naturally an $\mathfrak{su}(1,1)$-valued one form. $F[A]$ is then its curvature. $\phi$ is the scalar field. $\alpha$, $\Lambda$ and $m$ are coupling constants. $\alpha$ contains the gravity coupling constant $G$ and is, up to numerical factors $\frac{1}{G}$. $\Lambda$ is the cosmological constant and $m$ is the mass of the field. Finally, $\star$ is the Hodge dual associated to the metric constructed out of the triad. We will choose the signature $(-\ +\ +\ +)$, which goes with the sign in front of the mass term. There is a slight subtlety here. Normally, if $g$ is the metric and $\omega$ is a $p$-form, then:
\begin{equation}
(\star \omega)_{\mu_1 ... \mu_{n-p}} = \frac{1}{p! \sqrt{|\det g|}} \omega_{\nu_1 ... \nu_p} \epsilon^{\nu_1 ... \nu_p \rho_1 ... \rho_{n-p}} g_{\mu_1 \rho_1} ... g_{\mu_{n-p} \rho_{n-p}}.
\end{equation}
$\epsilon^{\mu\nu\rho}$ is not a tensor here and is simply the Levi-Civita symbol (it is a tensor multiplied by a density). Namely, $\epsilon^{012} = 1$ and all the other terms can be deduced by full anti-symmetry. But we have used the first order expression for the action which uses $\det e$ and not the square-root of the determinant of the metric, which are equal only up to a sign. Here, we will rather use the following expression, which also solves the sign problem:
\begin{equation}
(\star \omega)_{\mu_1 ... \mu_{n-p}} = \frac{1}{p! (\det e)} \omega_{\nu_1 ... \nu_p} \epsilon^{\nu_1 ... \nu_p \rho_1 ... \rho_{n-p}} g_{\mu_1 \rho_1} ... g_{\mu_{n-p} \rho_{n-p}}.
\end{equation}

As we discussed, one can hope that this theory is exactly quantizable (or at least in some special cases like $m = 0$). It is however rather difficult because of a few road-blocks:
\begin{itemize}
	\item The gauge group is non-abelian. This leads to various difficulties when constructing well-defined version of operators.
	\item The classical theory is not always solvable. For instance, a simple homogeneous scalar field coupled to 3d quantum gravity does not have an exact solution linking the volume of the universe to the value of the field. Though this is not an argument against the existence of a quantum version of the model exists, it is a noteworthy difficulty.
	\item Even in the classical case of point particles coupled to 3d gravity, the exact solution is rather difficult to implement and involves a lot of book-keeping. \cite{tHooft:1992izc}
\end{itemize}
The main idea of this paper is then to study a simpler model. We will study a scalar field coupled to \textit{linear} gravity. This model is taken from Lee Smolin work \cite{Smolin:1992wj}. It can be understood as a limit $G \rightarrow 0$ (that is $\alpha \rightarrow \infty$) of usual gravity with the additional constraint that $\frac{A}{G}$ (or $\alpha A$) is constant. This leads to the following (detailed) action:
\begin{eqnarray}
S[e,A,\phi] &=& \int_{\mathcal{S}} \Big[ \frac{\alpha}{2} \epsilon_{IJK} \epsilon^{\mu\nu\rho} e_\mu^I (\partial_\nu A_\rho^{JK} - \partial_\rho A_\nu^{JK}) + \frac{\Lambda}{6} \epsilon_{IJK} \epsilon^{\mu\nu\rho} e_\mu^I e_\nu^J e_\rho^K \nonumber \\
&-& \frac{1}{12} \epsilon_{IJK} \epsilon^{\mu\nu\rho} e_\mu^I e_\nu^J e_\rho^K \left( e^\sigma_M e^\tau_N \eta^{MN} \right) \partial_\sigma \phi \partial_\tau \phi - \frac{m^2}{12} \epsilon_{IJK} \epsilon^{\mu\nu\rho} e_\mu^I e_\nu^J e_\rho^K \phi^2 \Big] \mathrm{d}^3 x .
\end{eqnarray}
In this writing, $\epsilon_{IJK}$ is the standard Levi-Civita symbol. $\epsilon^{\mu\nu\rho}$ is not a tensor though, and follows the same convention as the one we used for defining the Hodge star. Also, we have used the standard notation of $e^\mu_I$ to write the inverse of the triad.

In practice, we see that this amounts to removing the non-abelian term from the curvature of $A$. Everything else is left untouched. This theory is particularly interesting because, while still diffeomorphism invariant, with some natural constraints, it is equivalent to the free scalar field. Indeed, assuming that $\mathcal{S} \simeq \mathbb{R}^3$, that the various fields behave properly at infinity (vanish quickly at infinity with their derivatives or converge at infinity for the triad), and that the triad is invertible everywhere (which we have more or less assumed when writing its inverse), then we can solve the equations of motion. They are:
\begin{itemize}
	\item $\mathrm{d} e^I = 0$ for all $I$. This means that, since $\mathcal{S}$ is simply connected, there is a collection of fields $\Psi^I$ such that $e^I = \mathrm{d} \Psi^I$.
	\item The usual equation of motion for the scalar field on a curved background: $\star \mathrm{d} (\star \mathrm{d} \phi) - m^2 \phi = 0$.
	\item For $A$, we get:
	\begin{equation}
	\frac{\alpha}{2} \epsilon_{IJK} \epsilon^{\mu\nu\rho} F_{\nu\rho}^{JK}[A] + (\det e) \Lambda e^\mu_I = (\det e)\left[ \frac{1}{2} e^\mu_I \left( g^{\sigma\tau} \partial_\sigma \phi \partial_\tau \phi + m^2 \phi^2 \right) - e^\sigma_I \partial_\sigma \phi \partial^\mu \phi \right].
	\end{equation}
	This equation always has a solution as long as the right term has a vanishing divergence, which is just the conservation of energy.
\end{itemize}
We see then, that $A$ is completely fixed by the rest of the fields, that the equation on $\phi$ are correct as soon as we can show that the space is flat. This is actually not always true. Indeed, all we have is: $e^I = \mathrm{d} \Psi^I$ and $e$ is invertible. This translates to $\epsilon_{IJK} \mathrm{d}\Psi^I \wedge \mathrm{d}\Psi^J \wedge \mathrm{d}\Psi^K \neq 0$ which means that the transformation from $\mathcal{S}$ to $\mathbb{R}^3$ encoded by $\Psi$ is \textit{locally} invertible. This sadly does not imply global invertibility. It should be noted however that this is part of the space of solutions. And when it is globally invertible, then is true that space is flat and we get the standard free field theory.

So we still get something interesting: the free scalar field is an entire sector of our theory. At this stage, it is quite unclear if this sector can be quantized independently from the others, but it is surely a fair assumption. We have a theory, therefore, that is diffeomorphism invariant and still contains the free scalar field. We should notice here similarities with parametrized field theory (PFT) \cite{Kuchar:1989bk,Kuchar:1989wz,Varadarajan:2006am}. And indeed, working with PFT really corresponds to directly working with $\Psi^I$. Compared to PFT, in addition to using directly the triad, we will also develop new directions for quantizing such a theory.

As the goal at this point is to write the corresponding quantum theory, we should be able to find quantities more or less equivalent to the creation and annihilation operators in standard quantum field theory. Indeed, if the free scalar field is an entire sector of the theory, this sector should be in correspondence with the usual solutions. We expect in particular corresponding ladder operators acting in this sector, though these quantities should probably be amended to accommodate the new symmetries.

What do we expect? A nice way to look at this is to consider an even simpler theory. Let's study a simple harmonic oscillator, that we can describe by the following action:
\begin{equation}
S = \int \left(\frac{1}{2}m\dot{x}^2 - \frac{1}{2}kx^2\right) \mathrm{d}t .
\end{equation}
Let's write this in a Hamiltonian manner. The momentum is:
\begin{equation}
p = m\dot{x} .
\end{equation}
This leads to the following Hamiltonian:
\begin{equation}
H = \frac{p^2}{2m} + \frac{kx^2}{2}.
\end{equation}
If we define $\omega = \sqrt{\frac{k}{m}}$, we can now write:
\begin{equation}
H = \frac{p^2}{2m} + m \frac{\omega^2 x^2}{2}.
\end{equation}
Now let's define the complex quantity:
\begin{equation}
a = \sqrt{\frac{m\omega}{2}} x  + \mathrm{i} \frac{p}{\sqrt{2m\omega}}
\end{equation}
And we finally have:
\begin{equation}
H = \omega a \overline{a} .
\end{equation}
It is now well-known that $a$ and $\overline{a}$ becomes creation and annihilation operators in the quantum theory.

Let's now turn to a diffeomorphism invariant version of this problem, starting with:
\begin{equation}
S = \int \left(\frac{1}{2}m\frac{\dot{x}^2}{\dot{t}} - \frac{1}{2}kx^2 \dot{t}\right) \mathrm{d}s ,
\end{equation}
where now $t$ is a variable depending on the parameter $s$ and all derivatives are taken with respect to $s$. A reparametrization will leave the action invariant which is therefore promoted to a diffeomorphism invariant one. We now have two momenta $p_x$ and $p_t$. And a complete Hamiltonian analysis will reveal that they must now satisfy a (first class) constraint which is:
\begin{equation}
p_t + \frac{p_x^2}{2m} + \frac{kx^2}{2} = 0 ,
\end{equation}
which is quite unsurprisingly the Shcrödinger equation (in its classical form). The interesting question though is can we adapt the $a$ quantity so that it commutes with this constraint?

Yes we can. The commutator of the current $a$ and our constraint is nearly zero already. In fact, the commutator with $p_t$ is zero but there is a constant (which is just the quanta of energy) for the second part. We must therefore add a term that does not commute with $p_t$. There are various ways to do that. The most interesting to us, is to just consider the time dependent expression for $a$. Indeed, $a$ follows the following equation of motion:
\begin{equation}
\frac{\mathrm{d}a}{\mathrm{d}t} = - \mathrm{i}\omega a .
\end{equation}
As a consequence:
\begin{equation}
a(t) = \left( \sqrt{\frac{m\omega}{2}} x  + \mathrm{i} \frac{p}{\sqrt{2m\omega}} \right) \mathrm{e}^{-\mathrm{i}\omega t} .
\end{equation}
Taken without modification, and by interpreting the $t$ as the conjugate to $p_t$, this quantity directly commutes with the constraint. This observation is what motivates our construction for the full system.

Our goal will be to reexpress the usual creation and annihilation operators in standard quantum field theory, so that the quantities linked to position and time can be reinterpreted in function of our new variables (the triad and the connection). If such a quantity can be constructed, it is by definition equal to the creation and annihilation operators when the gauge is fixed. But if it also commutes with the constraints, as our small study suggests, then it is a gauge-unfixed version of these operators and are really the natural operators in the diffeomorphism invariant world.

What we need to do then, is to get the Hamiltonian version of our problem. Then we will need to extract all the interesting operators as we just illustrated. This is what we do in the next section.

\section{Classical model}

\subsection{Hamiltonian analysis}

Ok, we now have the action we want to study. Let's start the Hamiltonian analysis proper. There are various mathematical difficulties we will just ignore for now. Namely, there are questions surrounding the behaviour of the fields at infinity or the various possible topologies for $\mathcal{S}$ the spacetime manifold. We will concentrate on the simplest possibility. All the other possibilities will just create a richer theory for which we will have neglected various sectors.

We will assume that $\mathcal{S}$ is homeomorphic to $\mathbb{R}^3$. We will also assume that all the matter fields vanish at infinity. Granted all this, we choose some decomposition of $\mathcal{S}$ as $\mathbb{R}\times\Sigma$ with corresponding coordinates $(t,\sigma)$. $t$ will be our time variable and $\sigma$ will be the coordinates on the spatial slice $\Sigma$. We do assume that $\Sigma$ is homeomorphic (and even diffeomorphic) to $\mathbb{R}^2$ but not necessarily a flat slice though. We also make the strong assumption that $\Sigma$ is spacelike with respect to the metric and nowhere degenerate. This last assumption is reasonable though as, in a hamiltonian analysis, we are interested in parametrizing the space of solutions which should correspond to the variables on a Cauchy slice of spacetime.

This allows the following writing:
\begin{equation}
S[e,A,\phi] = \int_\mathbb{R} L \mathrm{d}t,
\end{equation}
with:
\begin{eqnarray}
L &=& \int_{\Sigma} \Big[ \frac{\alpha}{2} \epsilon_{IJK} \epsilon^{\mu\nu\rho} e_\mu^I (\partial_\nu A_\rho^{JK} - \partial_\rho A_\nu^{JK})+ \frac{\Lambda}{6} \epsilon_{IJK} \epsilon^{\mu\nu\rho} e_\mu^I e_\nu^J e_\rho^K \nonumber \\
&-& \frac{1}{12} \epsilon_{IJK} \epsilon^{\mu\nu\rho} e_\mu^I e_\nu^J e_\rho^K \left( e^\sigma_M e^\tau_N \eta^{MN} \right) \partial_\sigma \phi \partial_\tau \phi - \frac{m^2}{12} \epsilon_{IJK} \epsilon^{\mu\nu\rho} e_\mu^I e_\nu^J e_\rho^K \phi^2 \Big] \mathrm{d}^2 \sigma .
\end{eqnarray}
From there, we proceed as usual: define the momenta, reverse the expressions that can be, keep the rest as primary constraints. The details of the computation can be found in appendix \ref{app:hamil}. Once all this is done, we can write the Legendre transform of the Lagrangian which is the Hamiltonian.

After some computations (detailed in the appendix), we finally get:
\begin{eqnarray}
H &=& \int_{\Sigma} \Big[ \frac{1}{2} \partial_0 A_0^{IJ} B^0_{IJ} + \frac{1}{2} \partial_0 A_a^{IJ} \left( B^a_{IJ}  - 2 \alpha \epsilon_{IJK} \epsilon^{ab} e_b^K\right) + X^\mu_I \partial_0 e_\mu^I - \frac{1}{2} A_0^{JK} \left(- 2 \alpha \epsilon_{IJK} \epsilon^{ab} \partial_b e_a ^I \right)  \nonumber \\
&-& e_0^I \Big(\alpha \epsilon_{IJK} \epsilon^{ab} F_{ab}^{JK}[A] + \Lambda n_I - \frac{1}{2} n_I h^{cd} \partial_c \phi \partial_d \phi - \frac{m^2}{2} n_I \phi^2 - \frac{n_I}{2 \det h} \Pi^2 \nonumber \\
&-& \frac{n_J \eta^{JK} \epsilon^{cd} \epsilon_{IKL} e_d^L}{\det h} \Pi \partial_c \phi \Big) \Big] \mathrm{d}^2 \sigma ,
\end{eqnarray}
with the following primary constraints:
\begin{equation}
\left\{\begin{array}{rcl}
X^0_I &=& 0, \\
B^0_{IJ} &=& 0, \\
X^a_I &=& 0, \\
B^a_{IJ} &=& 2\alpha \epsilon_{IJK} \epsilon^{ab} e_b^K.
\end{array}\right.
\end{equation}
Here, summations on small latin indices cover only spatial coordinates. Capital latin indices do cover the $3$ dimensions. $X$ is the natural conjugate with respect to $e$, $B$ the conjugate with respect to $A$ and $\Pi$ the conjugate of $\phi$. We have also used the following notations in the Hamiltonian:
\begin{itemize}
	\item $h_{ab}$ is the induced metric on $\Sigma$ and can be written as $h_{ab} = e_a^I e_b^J \eta^{IJ}$. Due to our assumptions, it is spacelike. $h^{ab}$ is the corresponding inverse metric.
	\item $n_I$ is the natural normal to $\Sigma$. It is a vector valued density and reads: $n_I = \frac{1}{2} \epsilon_{IJK} \epsilon^{ab} e_a^J e_b^K$.
\end{itemize}

From there, we can pursue the constraint analysis. After some lengthy, but straightforward, computations (see appendix \ref{app:hamil}), we get the following system of constraints:
\begin{equation}
\left\{\begin{array}{rcl}
0 &=& X^0_I, \\
0 &=& B^0_{IJ}, \\
0 &=& X^a_I, \\
0 &=& B^a_{IJ} - 2\alpha \epsilon_{IJK} \epsilon^{ab} e_b^K, \\
0 &=& -\alpha \epsilon_{IJK} \epsilon^{ab} F_{ab}^{JK}[A] - \Lambda n_I + \frac{1}{2} n_I h^{cd} \partial_c \phi \partial_d \phi + \frac{m^2}{2} n_I \phi^2 + \frac{n_I}{2 \det h} \Pi^2 + \frac{n_J \eta^{JK} \epsilon^{cd} \epsilon_{IKL} e_d^L}{\det h} \Pi \partial_c \phi, \\
0 &=& 2 \alpha \epsilon_{IJK} \epsilon^{ab} \partial_b e_a^I.
\end{array}\right.
\end{equation}
It can then be separated into first and second class constraints. We get two sets of second class constraints which are the equivalent of the simplicity constraints in 3d \cite{Charles:2017srg}:
\begin{equation}
\left\{\begin{array}{rcl}
0 &=& X^a_I, \\
0 &=& B^a_{IJ} - 2\alpha \epsilon_{IJK} \epsilon^{ab} e_b^K.
\end{array}\right.
\end{equation}
And we get a system of first class constraints:
\begin{equation}
\left\{\begin{array}{rcl}
0 &=& X^0_I, \\
0 &=& B^0_{IJ}, \\
0 &=& \partial_b B^b_{IJ}, \\
0 &=& \alpha \epsilon_{IJK} \epsilon^{ab} F_{ab}^{JK}[A] + \Lambda \tilde{n}_I - \frac{1}{2} \tilde{n}_I \tilde{h}^{cd} \partial_c \phi \partial_d \phi - \frac{m^2}{2} \tilde{n}_I \phi^2 - \frac{\tilde{n}_I}{2 \det \tilde{h}} \Pi^2 - \frac{\tilde{n}_J \eta^{JK} \epsilon^{cd} \epsilon_{IKL} \tilde{e}_d^L}{\det \tilde{h}} \Pi \partial_c \phi. 
\end{array}\right.
\end{equation}
where the tilded quantitites are constructed out of $B$ rather than $e$.

This allows the computation of the Dirac brackets:
\begin{equation}
\left\{\begin{array}{rcl}
\{e^I_0(x), X_J^0(y)\}_D &=& -\delta^I_J \delta(x-y),\\
\{A^{IJ}_0(x), B_{KL}^0(y)\}_D &=& -(\delta^I_K \delta^J_L - \delta^I_L \delta^J_K) \delta(x-y),\\
\{A^{IJ}_a(x), e^{K}_b(y)\}_D &=& \frac{1}{2\alpha \det h} \epsilon_{ab} \epsilon^{IJK} \delta(x-y),\\
\{A^{IJ}_a(x), B_{KL}^b(y)\}_D &=& -\delta_a^b (\delta^I_K \delta^J_L - \delta^I_L \delta^J_K) \delta(x-y),\\
\{\phi(x), \Pi(y)\}_D &=& -\delta(x-y),
\end{array}\right.
\end{equation}
all other (non-fundamental) brackets being zero (including brackets dealing with $X_I^a$). With these brackets, it is rather obvious that the second class constraints commute with all the other constraints. Interestingly, they can be solved, and the system can finally be rewritten as:
\begin{equation}
\left\{
\begin{array}{rcl}
0 &=& \alpha \epsilon_{IJK} \epsilon^{ab} F_{ab}^{JK}[A] + \Lambda n_I - \frac{1}{2} n_I h^{cd} \partial_c \phi \partial_d \phi - \frac{m^2}{2} n_I \phi^2 - \frac{n_I}{2 \det h} \Pi^2 - \frac{n_J \eta^{JK} \epsilon^{cd} \epsilon_{IKL} e_d^L}{\det h} \Pi \partial_c \phi, \\
0 &=& \epsilon^{ab} \partial_b e_a^I,
\end{array}
\right.
\end{equation}
with the following brackets:
\begin{equation}
\left\{\begin{array}{rcl}
\{A^{IJ}_a(x), e^{K}_b(y)\} &=& \frac{1}{2\alpha \det h} \epsilon_{ab} \epsilon^{IJK} \delta(x-y),\\
\{\phi(x), \Pi(y)\} &=& -\delta(x-y).
\end{array}\right.
\end{equation}
The $B$ variables have been removed thanks to the second class constraints and the time component variables have been removed as they decouple from the rest and can be trivially solved. We now have the Hamiltonian formulation of our problem.

How is this theory supposed to be linked to the free field theory? It is quite obvious that the constraint on the triad really carries the information that space is flat. There are a few subtleties linked to the problem of global invertibility we mentionned earlier but appart from this, it should be interpreted as the fact that the integral of $e$ is a vector that embed of surface $\Sigma$ into $\mathbb{R}^3$. The second constraint is familiar in its form (it is really the Einstein equation) but only set the value of the spin connection $A$. Apart from topological obstructions (which we avoided by choosing the simplest case), this equation always has a solution. So, where is the dynamics of the field encoded?

The point we have to remember is that the dynamics do not impose anything on a given Cauchy surface. As a consequence, $\phi$ and $\Pi$ are completely free. The only constraint will come from the evolution in time which should be encoded here as an action of the diffeomorphism constraints (they can be constructed out of the Einstein equation by projecting using $e$ and $n$). Therefore, the dynamics is not encoded in a constraint \textit{per se} but rather in their action. The constraint must be contained in the brackets with the curvature constraints. Because the equivalence has been established using the equations of motion earlier, we won't dwell into the equivalence here, which would require a careful analysis of possible gauge fixation. Rather, we will admit that this Hamiltonian theory should at least contain the free field theory and try from there to construct interesting quantities. In particular, we will study in the next section if it is possible to construct the equivalent of the creation and annihilation operators.

\subsection{Creation and annihilation operators}
\label{sec:basic_ops}

So we are looking for operators that should reduce in the correct gauge fixing to the standard creation and annihilation operator for the scalar field. In the diffeomorphism invariant context though, we expect them to commute with the constraints but still preserve a nice algebra among them, as was suggested on our simple harmonic oscillator study.

The difficulty resides in that the space manifold $\Sigma$ is not necessarily flat. The expression must therefore be adapted. We can go about two methods of construction. A first method would be to take advantage of the fact that $\Sigma$, though not flat, is supposed to be a Cauchy surface. This means that the field in the entire spacetime can be reconstructed from $\Pi$ and $\phi$ on the surface. The creation and annihilation operators could then be deduced as coefficient of the Fourier transform. This method would actually work (and it will be explored in section \ref{subsec:Fourier} to prove a couple of interesting properties) but is more complicated than necessary for now. A second idea is just to make a simple ansatz and check that the resulting operators have the correct algebra, among themselves but also with the constraints.

Let's go back to the standard free field theory for a moment. We have the following action:
\begin{equation}
S = -\int \frac{1}{2} \left(\eta^{\mu \nu} \partial_\mu \phi \partial_\nu \phi + m^2 \phi^2\right) \mathrm{d}^2 x \mathrm{d} t.
\end{equation}
This action leads to the following Hamiltonian:
\begin{equation}
H = \frac{1}{2} \int \left( \Pi^2 + (\vec{\nabla} \phi)^2 + m^2\phi^2 \right) \mathrm{d}^2 x ,
\end{equation}
where, once again $\Pi$ is conjugate to $\phi$. Normally, we define:
\begin{equation}
a_{\vec{k}} = \frac{1}{\sqrt{4\pi\omega_{\vec{k}}}}\int \left(\omega_{\vec{k}} \phi  + \mathrm{i} \Pi \right) \exp\left( -\mathrm{i} \vec{k}\cdot \vec{x} \right) \mathrm{d}^2 x ,
\end{equation}
where $\omega_{\vec{k}} = \sqrt{\vec{k}^2 + m^2}$. This allows the simple expression:
\begin{equation}
H = \int \omega_{\vec{k}} \overline{a_{\vec{k}}} a_{\vec{k}} \mathrm{d}^2 k .
\end{equation}
And of course, we have the well-known algebra:
\begin{equation}
\left\{\begin{array}{rcl}

\{a_k, \overline{a_{k'}}\} &=& \mathrm{i} \delta(k-k') , \\
\{H, a_k\} &=& -\mathrm{i} \omega_k a_k , \\
\{H, \overline{a_k}\} &=& \mathrm{i} \omega_k \overline{a_k} .

\end{array}\right.
\end{equation}

Can we have a similar algebra with the coupling to linear gravity? The problem comes from the Hamiltonian which no longer exists but is replaced by a collection of constraints. The curvature constraints (which contain the Einstein equation projected on $\Sigma$) are however local. We can show the problem with this in the non-gravitational case, by looking at the commutator not with the Hamiltonian $H$ but rather with $H(x) = \frac{1}{2}\left(\Pi^2 + (\vec{\nabla} \phi)^2 + m^2\phi^2\right)$ which is the integrand. We get:
\begin{equation}
\{H(x), a_k\} = \frac{1}{\sqrt{4\pi\omega_k}} \left(- \vec{\nabla}\phi \cdot \vec{k} - \mathrm{i} m^2 \phi + \omega_k \Pi\right) \exp\left( -\mathrm{i} \vec{k}\cdot \vec{x} \right) .
\end{equation}
The resulting expression is not integrated over space, depends on the derivatives of $\phi$ and cannot simply be expressed in terms of the creation and annihilation operators. How can we solve these problems?

What must happen is similar to what we have seen in the case of the harmonic oscillator: the curvature of  $A$ in the curvature constraint will not commute with the operators and will exactly compensate. This is possible if some part of the creation-annihilation operators uses the triad. The natural way to do this, is to use the integral of the triad as a position operator.

So, let's start from this kind of expressions:
\begin{equation}
a_k = \int \left(f(k,\sigma,e,A) \phi + g(k,\sigma,e,A) \Pi\right)\mathrm{d}^2\sigma .
\end{equation}
This is just the most generic linear expression. Can we go further? Well somewhat yes. We want two additionnal properties:
\begin{enumerate}
	\item The expression should be covariant with respect to local gauge transforms.
	\item The expression should be covariant (or even invariant) with respect to diffeomorphism transforms.
\end{enumerate}
Concerning the first point, we do expect some covariance. Basically, $k$ should be expressed in some local reference frame and when it is changed, $k$ should change meaning some covariance for $a_k$. In the linear gravity scenario though, the reference frames cannot change by gauge transform (an interpretation of this is that only infinitesimal changes have been kept). We therefore expect full invariance. This leads to the simple condition that $a_k$ should commute with the Gauß constraint ($\mathrm{d}e = 0$). As $e$ is invariant under Gauß transforms, then this means that $a_k$ can depend on $A$ only through its curvature.

Something similar can be said for diffeomorphism invariance. In principle, in the full theory, we only expect some kind of covariance. One problem for instance is that the integral of (parallel transported) $e$ depends on the path and so the annihilation operator could be linked to some integration path choice. In that case, diffeomorphism transform might lead to some transformation of the operators. We are in the linear gravity case though. And in that case, it is way easier to solve. The integral of $e$ does not depend on the choice of path (thanks to the Gauß constraint). So we can make similarly the reasonnable assumption that $a_k$ should be invariant under diffeomorphism transforms.

This leads to the following expression:
\begin{equation}
a_k = \int \left(\tilde{f}(k,\sigma,e,F[A]) \phi + \tilde{g}(k,\sigma,e,F[A])\Pi\right)\mathrm{d}^2\sigma .
\end{equation}
with the additional constraint that $a_k$ commutes with the curvature constraints. We can make one additional assumption: that $a_k$ does not depend on $A$ at all. This seems reasonable enough since we don't really see how this would enter the equation anyway and the standard creation operator doesn't have any dependence on curvature (at least for scalars).

So, we have the following working hypothesis. The annihilation operator has the following form:
\begin{equation}
a_k = \int \left(h_1(k,\sigma,e) \phi + h_2(k,\sigma,e)\Pi\right)\mathrm{d}^2\sigma .
\end{equation}
And:
\begin{equation}
\{D_I, a_k\}_D = 0.
\end{equation}
A nice addition is to use our guess about the depency in the triad for the position operators.. We offer the following ansatz:
\begin{equation}
a_k = \frac{1}{\sqrt{2}\pi}\int \left(A(k,e,\sigma) k^I n_I \phi + \mathrm{i} B(k,e,\sigma) \Pi\right) \mathrm{e}^{- \mathrm{i} \vec{k} \cdot \int^{\sigma} \vec{e}} \mathrm{d}^2\sigma .
\end{equation}
This expression is directly inspired from the standard expression for the annihilation operator. Let's explain a few bits:
\begin{itemize}
	\item The factor $k^I n_I$ is a density. This way $A$ is a scalar. It might not be the right density to put (for instance $\sqrt{n^I n_I}$ would work too) but this doesn't matter since it can be corrected with the right expression for $A$ (which would then be the ratio between two densities). It is a natural\footnote{There are other possibilities that reflect this though: for instance $k^I n_I(Q) \sqrt{n^I n_I}$ where $Q$ is some fixed reference point on the manifold. But once more, this can be done by adjusting $A$, thought this might be taken as some explicit dependancy on $\sigma$. So let's not forget this possibility later on.} density to consider though since it very much looks like the energy component of $k$.
	\item The integral term $\int^{\sigma} \vec{e}$ is a bit weird to say the least. First, $\vec{e}$ is simply the triad taken to be a vector-valued one-form. Now the integral only has an end point of coordinates $\sigma$. But the fact that there is no start point is actually important: we \textit{cannot} take a specific point as reference. Indeed, the exponential of the triad creates curvature at one point and destroys it at the other. Here, we need an operator that only create curvatures at a specific point.
	
	This operator really corresponds to the $\Psi$ we encountered earlier such that $\mathrm{d}\Psi^I = e^I$. Because of this relation ship with the triad, there is still a sense in which the difference of $2$ $\Psi$ is an integral of the triad. By extension, we use this notation with only one end-point to the integral.
	
	There is a way to make this more rigorous for a non-compact spatial slice. Because, all the information is contained in a Dirac bracket, we can consider the action of the integral as the start points goes to infinity. Though the integral is not well-defined, its Dirac bracket still exists and correspond exactly to what we need.
\end{itemize}
It turns out that the correct values are:
\begin{equation}
\left\{
\begin{array}{rcl}
A(k,e,\sigma) &=& 1, \\
B(k,e,\sigma) &=& 1.
\end{array}
\right.
\end{equation}
This leads to the following, and in fact quite familiar, expression:
\begin{equation}
a_k = \frac{1}{\sqrt{2}\pi}\int \left(k^I n_I \phi + \mathrm{i}\Pi\right) \mathrm{e}^{- \mathrm{i} \vec{k} \cdot \int^{\sigma} \vec{e}} \mathrm{d}^2\sigma .
\end{equation}
A lengthy - but not difficult - computation shows that indeed (see appendix \ref{app:operators}):
\begin{equation}
\{D_I, a_k\}_D = 0.
\end{equation}
More interestingly, the algebra of these operators can be computed explicitly. It requires some technology we will develop in the next section.

\subsection{Fourier transform and full algebra}

\label{subsec:Fourier}

A point must be underlined here: in usual free field theory, the creation and annihilation operators have a nice interpretation as Fourier coefficients of the 3d field solution of the equation of motion. A similar property holds true here, granted a few assumptions.

Our spacetime is $\mathbb{R}^3$ (this was one of our simplifying assumptions). We also assumed that $\Sigma$ (the space manifold) is homeomorphic to $\mathbb{R}^2$. We will go a bit further here and assume that the embedding of $\Sigma$ into $\mathbb{R}^3$ given by the integrals of the triads $\int \vec{e}$ is a Cauchy surface for the free field theory. This assumption is reasonable: when we choose a slice $\Sigma$ of spacetime, our goal is not to break diffeomorphism invariance but to parameterize the space of solutions for the problem. It is natural therefore to choose a Cauchy surface to do so. It is even natural to think that if we don't choose a Cauchy surface, the Hamiltonian analysis will not be well-defined. We will leave this question open however and just assume a correct choice of $\Sigma$.

What we mean by this assumption is the following. Let $\phi : \mathbb{R}^3 \mapsto \mathbb{R}$ be a field that satisfies the standard free scalar field equation:
\begin{equation}
-\partial_t^2 \phi + \Delta \phi - m^2 \phi = 0.
\end{equation}
Let's now interpret $\Sigma$ as a submanifold of $\mathbb{R}^3$ with embedding given by $\vec{\Psi} = \int \vec{e}$. We assume that knowing $\phi$ and its derivative along the normal on this embedding is sufficient (and also necessary) to know $\phi$ on the whole $\mathbb{R}^3$. This means that we can now extend naturally some fields on $\Sigma$ to the whole $\mathbb{R}^3$ spacetime.

On the $\Sigma$ slice, we have two fields we are interested in $\phi$ and $\Pi$. $\Pi$ can naturally be connected to a derivative of $\phi$ in the time-direction (see appendix \ref{app:hamil}):
\begin{equation}
\Pi = -(\det e)g^{0\tau} \partial_\tau \phi = -\vec{n} \cdot \vec{\nabla} \phi.
\end{equation}
Here, $\vec{n}$ is the normal density on $\Sigma$ induced by the triad and $\vec{\nabla} \phi$ is the gradient of $\phi$ (as a spacetime field) expressed in the coordinates we used for the embedding. This means that $\phi$ and $\Pi$ on $\Sigma$ can naturally be extended to a field on the whole spacetime $\mathbb{R}^3$. Now, we can use the Fourier transform as usual on $\mathbb{R}^3$ and get coefficients that will turn out to be the $a_k$ we defined earlier (up to some Dirac deltas factor). But of course, the formula will be more general and apply to any couple of fields we might define on $\Sigma$.

Now, let's turn back to our expression for $a_k$:
\begin{equation}
a_k = \frac{1}{\sqrt{2}\pi}\int \left(k^I n_I \phi + \mathrm{i}\Pi\right) \mathrm{e}^{- \mathrm{i} \vec{k} \cdot \int^{\sigma} \vec{e}} \mathrm{d}^2\sigma .
\end{equation}
Our claim is that, this is (up to a factor we will make explicit shortly) the Fourier coefficients for the extension of $\phi$ in $\mathbb{R}^3$ according to the previous rules. There is a rather simple way to check this thanks to linearity. We just have to consider the case of:
\begin{equation}
\left\{
\begin{array}{rcl}
\phi(\sigma) &=& A\frac{\delta(\sigma-\sigma_0)}{\sqrt{\det h}}, \\
\Pi(\sigma) &=& B\delta(\sigma - \sigma_0).
\end{array}
\right.
\label{eq:ini}
\end{equation}
We have put the determinant for $\phi$, because $\phi$ is a scalar and we want $A$ not to depend on the choice of coordinates. $\Pi$ however is a density, and so to have $B$ coordinate independent, the determinant factor should be avoided. In that case:
\begin{equation}
a_k = \frac{1}{\sqrt{2}\pi} \left(\frac{k^I n_I(\sigma_0)}{\sqrt{\det h(\sigma_0)}} A + \mathrm{i}B\right) \mathrm{e}^{- \mathrm{i} \vec{k} \cdot \int^{\sigma_0} \vec{e}} .
\end{equation}
Let's now consider a field $\Phi(x,t)$ solution of the equation of motion in $\mathbb{R}^3$. We can write it in a general form as follows:
\begin{equation}
\Phi(\vec{x}) = \int \delta(k^2 + m^2)b_k \mathrm{e}^{\mathrm{i}\vec{k}\cdot \vec{x}} \mathrm{d}^3 k.
\end{equation}
The $b_k$ are therefore the Fourier coefficients (up to a Dirac delta factor) of $\Phi$. Let's now consider the plane $\mathcal{P}$ going through $\int^\sigma_0 \vec{e}$ and tangent to $\Sigma$ (or more precisely tangent to its embedding) at this point. This plane is spacelike and as such can be used as a Cauchy surface for the field $\Phi$.

There is always a Lorentz transformation sending $(1,0,0)$ to the normalized normal of the plane $\mathcal{P}$, granted the chosen orientation is the same (there is an infinite amount of such transformation but anyone will do, we can for instance take a boost). Let's note such a Lorentz transformation $L$. We can now write a parametrisation of the points of $\mathcal{P}$ as follows:
\begin{equation}
\vec{x}_\mathcal{P}(\tilde{X}) = \overrightarrow{L\triangleright(0,\tilde{X})}+\int^{\sigma_0} \vec{e}.
\end{equation}
Here we chose the following notation: to a vector $\vec{z}$ can be associated a 2d spatial vector $\tilde{z}$ and a time component $z_t$. By extension, any 2d vector will be written $\tilde{w}$ as we used for the coordinates on the plane denoted $\tilde{X}$. Also, $\triangleright$ is used to indicate the action of the Lorentz group onto 3d vectors. We can now write initial conditions on the plane $\mathcal{P}$ for $\Phi$:
\begin{equation}
\forall \tilde{X}\in\mathbb{R}^2,\ \left\{\begin{array}{rcl}
\Phi(\vec{x}_\mathcal{P}(\tilde{X})) &=& A\delta(\tilde{X}), \\
-\overrightarrow{L\triangleright(1,0,0)}\cdot\vec{\nabla}\Phi(\vec{x}_\mathcal{P}(\tilde{X})) &=& B\delta(\tilde{X}).
\end{array}
\right.
\end{equation}
These initial conditions correspond to the values of equation \ref{eq:ini}. Indeed, thanks to the Minkowski structure of spacetime, nothing can propagate faster than light. With the conditions of equation \ref{eq:ini}, this translates to $\Phi(x) = 0$ for any point outside of the lightcone of the point at $\sigma_0$. Now, the transformations laws under diffeomorphism are completely local which guarantees that $\Phi$ is a Dirac delta on any Cauchy surface passing through $\sigma_0$. The fact that $\Phi$ is a scalar even gives the coefficient of transformation which is $1$. We must however be careful, as the Dirac delta is a density, which is why the determinant is eaten up. A similar result holds for the derivative: it is zero nearly everywhere and locally can be expressed with respect to the gradient on $\Sigma$ and $\Pi$. Because, we chose a surface tangent to $\Sigma$, the gradient does not appear and we can conclude.

We can now use the standard derivation of $b_k$ in terms of $A$ and $B$. Let $\vec{k}$ be a 3d vector with $k^2 + m^2 = 0$ and $k_t > 0$. Then, we get:
\begin{equation}
\int \Phi(\vec{x}_\mathcal{P}(\tilde{X})) \mathrm{e}^{-\mathrm{i}\vec{k}\cdot \vec{x}_\mathcal{P}(\tilde{X})} \mathrm{d}^2 \tilde{X} = A\mathrm{e}^{-\mathrm{i}\vec{k}\cdot \vec{x}_\mathcal{P}(0)}.
\end{equation}
We can also compute:
\begin{eqnarray}
& & \int \Phi(\vec{x}_\mathcal{P}(\tilde{X})) \mathrm{e}^{-\mathrm{i}\vec{k}\cdot \vec{x}_\mathcal{P}(\tilde{X})} \mathrm{d}^2 \tilde{X} \nonumber \\
&=& \int \int \delta((k')^2 + m^2)b_{k'} \mathrm{e}^{\mathrm{i}\vec{k'}\cdot \vec{x}_\mathcal{P}(\tilde{X})} \mathrm{d}^3 k' \mathrm{e}^{-\mathrm{i}\vec{k}\cdot \vec{x}_\mathcal{P}(\tilde{X})} \mathrm{d}^2  \tilde{X} \nonumber \\
&=& \int \int \delta((k')^2 + m^2)b_{k'} \mathrm{e}^{\mathrm{i}\left(L^{-1} \triangleright (\vec{k'} - \vec{k})\right)\cdot \left(L^{-1} \triangleright\vec{x}_\mathcal{P}(\tilde{X})\right)} \mathrm{d}^3 k' \mathrm{d}^2  \tilde{X} \nonumber \\
&=& \int \int \delta((L\triangleright k')^2 + m^2)b_{L\triangleright k'} \mathrm{e}^{\mathrm{i}(\vec{k'} - L^{-1} \triangleright \vec{k})\cdot \left(L^{-1} \triangleright\vec{x}_\mathcal{P}(\tilde{X})\right)} \mathrm{d}^3 k' \mathrm{d}^2  \tilde{X} \nonumber \\
&=& \int \int \delta((k')^2 + m^2)b_{L\triangleright k'} \mathrm{e}^{\mathrm{i}(\tilde{k'} - (\tilde{L^{-1} \triangleright \vec{k}}))\cdot \tilde{X}} \mathrm{e}^{\mathrm{i}(\vec{k'} - L^{-1} \triangleright \vec{k})\cdot \left(L^{-1} \triangleright\vec{x}_\mathcal{P}(\tilde{0})\right)} \mathrm{d}^3 k' \mathrm{d}^2  \tilde{X} \nonumber \\
&=& (2\pi)^2 \int \delta((k')^2 + m^2)b_{L\triangleright k'}\delta(\tilde{k'} - (\tilde{L^{-1} \triangleright \vec{k}})) \mathrm{e}^{\mathrm{i}(\vec{k'} - L^{-1} \triangleright \vec{k})\cdot \left(L^{-1} \triangleright\vec{x}_\mathcal{P}(\tilde{0})\right)} \mathrm{d}^3 k' \nonumber \\
&=& (2\pi)^2 \int \frac{\delta\left(k'_t - \sqrt{\vec{k'}^2 + m^2}\right) + \delta\left(k'_t + \sqrt{\vec{k'}^2 + m^2}\right)}{2|k'_t|}b_{L\triangleright k'}\delta(\tilde{k'} - (\tilde{L^{-1} \triangleright \vec{k}})) \mathrm{e}^{\mathrm{i}(\vec{k'} - L^{-1} \triangleright \vec{k})\cdot \left(L^{-1} \triangleright\vec{x}_\mathcal{P}(\tilde{0})\right)} \mathrm{d}^3 k'.
\end{eqnarray}
This last line splits into two terms.
For the first line, the main observation is that:
\begin{equation}
\delta\left(k'_t - \sqrt{\vec{k'}^2 + m^2}\right)\delta(\tilde{k'} - (\tilde{L^{-1} \triangleright \vec{k}})) = \delta(\vec{k'} - L^{-1} \triangleright \vec{k})
\end{equation}
as there is a unique vector of square norm $-m^2$ with given spatial support and with positive time component. The second term is more involved. We get:
\begin{equation}
\delta\left(k'_t + \sqrt{(-\vec{k'})^2 + m^2}\right)\delta(\tilde{k'} - (\tilde{L^{-1} \triangleright \vec{k}})) = \delta(\vec{k'} - \overline{L^{-1} \triangleright \vec{k}}) ,
\end{equation}
where $\overline{\vec{x}}$ is the vector deduced from $\vec{x}$ by inverting its time component, namely $(-x_t, \tilde{x})$. This leads to:
\begin{eqnarray}
& & \int \Phi(\vec{x}_\mathcal{P}(\tilde{X})) \mathrm{e}^{-\mathrm{i}\vec{k}\cdot \vec{x}_\mathcal{P}(\tilde{X})} \mathrm{d}^2 \tilde{X} \nonumber \\
&=& (2\pi)^2 \int \frac{1}{2|k'_t|}\delta(\vec{k'} - L^{-1} \triangleright \vec{k})b_{L\triangleright k'} \mathrm{e}^{\mathrm{i}(\vec{k'} - L^{-1} \triangleright \vec{k})\cdot \left(L^{-1} \triangleright\vec{x}_\mathcal{P}(\tilde{0})\right)} \mathrm{d}^3 k' \nonumber \\
&+& (2\pi)^2 \int \frac{1}{2|k'_t|}\delta(\vec{k'} - \overline{L^{-1} \triangleright \vec{k}})b_{L\triangleright k'} \mathrm{e}^{\mathrm{i}(\vec{k'} - L^{-1} \triangleright \vec{k})\cdot \left(L^{-1} \triangleright\vec{x}_\mathcal{P}(\tilde{0})\right)} \mathrm{d}^3 k' \nonumber \\
&=& \frac{2\pi^2}{2(L\triangleright k)_t}\left( b_k + b_{\overline{k}}\mathrm{e}^{-2\mathrm{i} \left(L^{-1}\triangleright k\right)_t \left(L^{-1} \triangleright\vec{x}_\mathcal{P}(\tilde{0})\right)_t} \right)
\end{eqnarray}
Similarly, we can compute:
\begin{equation}
\int -\overrightarrow{L\triangleright(1,0,0)}\cdot\vec{\nabla}\Phi(\vec{x}_\mathcal{P}(\tilde{X})) \mathrm{e}^{-\mathrm{i}\vec{k}\cdot \vec{x}_\mathcal{P}(\tilde{X})} \mathrm{d}^2 \tilde{X} = B\mathrm{e}^{-\mathrm{i}\vec{k}\cdot \vec{x}_\mathcal{P}(0)},
\end{equation}
and also:
\begin{equation}
-\int \overrightarrow{L\triangleright(1,0,0)}\cdot\vec{\nabla}\Phi(\vec{x}_\mathcal{P}(\tilde{X})) \mathrm{e}^{-\mathrm{i}\vec{k}\cdot \vec{x}_\mathcal{P}(\tilde{X})} \mathrm{d}^2 \tilde{X} = -2\mathrm{i}\pi^2 (b_k - b_{\overline{k}}\mathrm{e}^{-2\mathrm{i} \left(L^{-1}\triangleright k\right)_t \left(L^{-1} \triangleright\vec{x}_\mathcal{P}(\tilde{0})\right)_t}).
\end{equation}
We can conclude:
\begin{equation}
\left\{
\begin{array}{rcl}
b_k &=& \frac{1}{2\pi^2}\left( (L^{-1}\triangleright k)_t A + iB \right)\mathrm{e}^{-\mathrm{i}\vec{k}\cdot \vec{x}_\mathcal{P}(0)}, \\
b_{\overline{k}} &=& \frac{1}{2\pi^2}\left( (L^{-1}\triangleright k)_t A - iB \right)\mathrm{e}^{-\mathrm{i}\overline{\vec{k}}\cdot \vec{x}_\mathcal{P}(0)}.
\end{array}
\right.
\end{equation}
Now:
\begin{eqnarray}
(L^{-1}\triangleright k)_t &=& (L^{-1}\triangleright \vec{k})\cdot\overrightarrow{(1,0,0,)} \nonumber \\
&=& \vec{k}\cdot(L^{-1}\triangleright \overrightarrow{(1,0,0,)}) \nonumber \\
&=& \frac{k^I n_I(\sigma_0)}{\sqrt{\det h(\sigma_0)}},
\end{eqnarray}
where we used $n$ divided by its norm as an expression for the normal to $\mathcal{P}$. Thus:
\begin{equation}
\left\{
\begin{array}{rcl}
b_k &=& \frac{1}{2\pi^2}\left( \frac{k^I n_I(\sigma_0)}{\sqrt{\det h(\sigma_0)}} A + iB \right)\mathrm{e}^{-\mathrm{i}\vec{k}\cdot \vec{x}_\mathcal{P}(0)}, \\
b_{\overline{k}} &=& \frac{1}{2\pi^2}\left( \frac{k^I n_I(\sigma_0)}{\sqrt{\det h(\sigma_0)}} A - iB \right)\mathrm{e}^{-\mathrm{i}\overline{\vec{k}}\cdot \vec{x}_\mathcal{P}(0)}.
\end{array}
\right.
\end{equation}
We can finally rewrite this in the more traditional manner:
\begin{equation}
\left\{
\begin{array}{rcl}
b_k &=& \frac{1}{2\pi^2}\left( \frac{k^I n_I(\sigma_0)}{\sqrt{\det h(\sigma_0)}} A + iB \right)\mathrm{e}^{-\mathrm{i}\vec{k}\cdot \vec{x}_\mathcal{P}(0)}, \\
b_{-k} &=& \frac{1}{2\pi^2}\left( -\frac{k^I n_I(\sigma_0)}{\sqrt{\det h(\sigma_0)}} A - iB \right)\mathrm{e}^{\mathrm{i}\vec{k}\cdot \vec{x}_\mathcal{P}(0)}.
\end{array}
\right.
\end{equation}
And then:
\begin{equation}
b_k = \frac{\sgn(k_t)}{\sqrt{2}\pi}a_k,
\end{equation}
and this is true for any $k$ such that $k^2 + m^2 = 0$.

All this means that, up to a numerical factor, the sign of $k_t$ and a Dirac delta, the $a_k$ coefficients really are the Fourier coefficients of the field we get by specifying the initial conditions of $\Phi$ and $\Pi$ on $\Sigma$ embedded into $\mathrm{R}^3$. This is especially useful to compute the brackets between the $a_k$ coefficients. Let's compute the following bracket:
\begin{eqnarray}
& & \{\delta(k^2 + m^2)a_k, \delta(k'^2 + m^2)a_{k'}\} \nonumber \\
&=& \delta(k^2 + m^2) \delta(k'^2 + m^2) \{a_k, a_{k'}\} \nonumber \\
&=& \delta(k^2 + m^2) \delta(k'^2 + m^2) \frac{1}{2\pi^2} \int \int \{k^I n_I(x) \phi(x) + \mathrm{i}\Pi(x), k'^J n_J(y) \phi(y) + \mathrm{i}\Pi(y)\} \mathrm{e}^{- \mathrm{i} \vec{k}\cdot \int^{x} \vec{e} - \mathrm{i}\vec{k'}\cdot \int^{y} \vec{e} } \mathrm{d}^2 x \mathrm{d}^2 y\nonumber \\
&=& \delta(k^2 + m^2) \delta(k'^2 + m^2) \frac{\mathrm{i}}{2\pi^2} \int \int \left( - k^I n_I(x) \delta(x-y) + k'^J n_J(y) \delta(x-y) \right) \mathrm{e}^{- \mathrm{i} \vec{k}\cdot \int^{x} \vec{e} - \mathrm{i}\vec{k'}\cdot \int^{y} \vec{e} } \mathrm{d}^2 x \mathrm{d}^2 y\nonumber \\
&=& \delta(k^2 + m^2) \delta(k'^2 + m^2) \frac{\mathrm{i}}{2\pi^2} \int (k' - k)^I n_I \mathrm{e}^{- \mathrm{i} (\vec{k} + \vec{k'})\cdot \int^{x} \vec{e}} \mathrm{d}^2 x.
\end{eqnarray}
Though this last form is pretty compact, it is better to expend it back a bit as follows:
\begin{eqnarray}
 & \{\delta(k^2 + m^2)a_k, \delta(k'^2 + m^2)a_{k'}\} = \nonumber \\
 & \delta(k'^2 + m^2) \left[ \delta(k^2 + m^2) \frac{1}{\sqrt{2}\pi} \int \left((-\frac{\mathrm{i}}{\sqrt{2}\pi} \mathrm{e}^{- \mathrm{i} \vec{k'}\cdot \int^{x} \vec{e}})k^I n_I + \mathrm{i} (\frac{1}{\sqrt{2}\pi}k'^I n_I \mathrm{e}^{- \mathrm{i} \vec{k'}\cdot \int^{x} \vec{e}})\right) \mathrm{e}^{- \mathrm{i} \vec{k}\cdot \int^{x} \vec{e}} \mathrm{d}^2 x \right]
\end{eqnarray}
From what we just saw, the term in large square brackets is (up to a numerical factor and a sign) the Fourier coefficient of a field with initial values on $\Sigma$ given by:
\begin{equation}
\left\{
\begin{array}{rcl}
\phi &=& -\frac{\mathrm{i}}{\sqrt{2}\pi} \mathrm{e}^{- \mathrm{i} \vec{k'}\cdot \int^{x} \vec{e}}, \\
\Pi &=& \frac{1}{\sqrt{2}\pi}k'^I n_I \mathrm{e}^{- \mathrm{i} \vec{k'}\cdot \int^{x} \vec{e}}. 
\end{array}
\right.
\end{equation}
But we know such a field: it is simply the field $\Phi(x) = -\frac{\mathrm{i}}{\sqrt{2}\pi} \mathrm{e}^{- \mathrm{i} k'\cdot x}$ on the whole $\mathbb{R}^3$ spacetime. And its Fourier transform is proportional to a Dirac delta $\delta(k+k')$. From that, we conclude (with the factors correctly computed):
\begin{equation}
\{\delta(k^2 + m^2)a_k, \delta(k'^2 + m^2)a_{k'}\} = -\mathrm{i}\sgn(k_t)\delta(k'^2 + m^2)\delta(k+k').
\end{equation}
This is exactly the kind of algebra we wanted for creation-annihilation operators. It is correctly adapted to the diffeomorphism invariant case as no frame of reference can be preferred. Let's note here that the sign is the reverse from the usual since we have:
\begin{equation}
\overline{a_k} = -a_{-k}
\end{equation}
with the extra sign coming from the fact that we put the $\sgn(k_t)$ factor out of $a_k$.

\section{Quantization}

\subsection{First approach}

\label{sec:firstapproach}

We can now turn to the quantization of the system. In principle, we should start with some natural construction of the algebra of observables, starting with canonical variables. This is however notoriously difficult for matter coupled to gravity \cite{Ashtekar:2002vh,Kaminski:2005nc,Kaminski:2006ta}. As a first approach, let's avoid the usual difficulties by choosing another set of fundamental variables.

The first point to note is that we have the creation and annihilation operators which are quite natural. They are for instance used in the construction of the Fock space and it does make sense to keep them as fundamental. The second point to note is that the creation and annihiliation operators, by construction, commute with the triad operators and with the curvature constraints. They commute with the triad because they do not depend on the connection, and we devoted a large part of this paper (see appendix \ref{app:operators}) to prove it commutes with the curvature constraints. Conversely, the triad operators and the curvature constraints are particularly interesting as fundamental variables since they are conjugate to each other. Finally, we have proven previously that the $a_k$ can be interpreted as Fourier coefficients (section \ref{subsec:Fourier}), which means we can reconstruct (at least classically) the field $phi$ and its momentum $\Pi$. This also means that, classically, if we now the triad and the curvature constraints, we can reconstruct the curvature of the connection everywhere. This is enough to reconstruct the spin connection up to a gauge. Therefore, the following collection:
\begin{itemize}
	\item $a_{k}$ for all $k \in \mathbb{R}^3$ such that $k^2 + m^2 = 0$ (which contains both creation and annihilation operators based on the sign of $k^0$),
	\item $D_I(x)$ for all $I$ and $x$,
	\item and $e_a^I(x)$ for all $I$, $a$ and $x$
\end{itemize}
gives a complete description of the gauge invariant phase-space. This collection divides into two sectors that commute with each other and that, remarkably, we know how to quantize separately. The creation-annihilation algebra leads to the well-known Fock quantization (with a few caveats). And the algebra of the curvature and triad operators can lead to a quantization around a state similar to the BF vacuum \cite{Dittrich:2014wpa,Bahr:2015bra} as we will shortly show.

There is one important point to underline here: all this works only when restricting to the gauge-invariant subspace of the phase space. It is not always possible to solve for this subspace explicitly, and it is not possible for the non-abelian case. In the abelian case however, not only is it possible, it greatly simplifies a number of expressions. Indeed, the algebra between the $D_I$ is only simple if the Gauß constraints is checked. The same thing holds for the brackets between $D_I$ and $a_k$ which in all generality is linear in the Gauß constraints. In general then, we would have to deal with partial gauge-fixing, the choice of path and other niceties. And such a treatment will be \textit{necessary} for the non-abelian case. However, as a first approach, and when considering our simple linear theory, it is possible to avoid such consideration. And this is what will do in all the constructions from now on.

\medskip

Let's start with the Fock quantization. We have shown that the creation-annihilation operators respect an algebra similar to the standard one. There is a caveat though, as this algebra is labeled by vectors in $\mathbb{R}^3$ (rather than $\mathbb{R}^2$) but with the additional constraint of being on the mass shell. This corresponds to functions living on the two-sheet hyperboloid, with the condition that reflection with respect to the origin gives rise to a complex conjugation.

If we want to map this algebra onto the usual one, we have to project these functions over the hyperboloid onto the plane $\mathbb{R}^2$. This can be done quite easily (though not in a covariant way) by considering only one sheet of the hyperboloid (the other one can be recovered by conjugation) and forgetting about the time component of the momentum $k$. For instance, let's restrict to the $k_t > 0$ sheet. We can define:
\begin{equation}
c_{\tilde{k}} = a_{(\sqrt{\tilde{k}^2 + m^2},\tilde{k})}.
\end{equation}
The $c$ operators now check an algebra that is even more familiar:
\begin{equation}
\left\{
\begin{array}{rcl}
\{c_{\tilde{k}}, c_{\tilde{k}'}\} &=& 0, \\
\{\overline{c_{\tilde{k}}}, \overline{c_{\tilde{k}'}}\} &=& 0, \\
\{c_{\tilde{k}}, \overline{c_{\tilde{k}'}}\} &=& 2\mathrm{i}\sqrt{\tilde{k}^2 + m^2} \delta(\tilde{k}-\tilde{k}').
\end{array}
\right.
\end{equation}
We notice here an energy factor. This is due to the unusual convention used for the $a$ as we did not divide by the square root of the energy. Though this was natural to preserve a covariant expression, this means that the square of $a$ operators (that is $N_k = a_k^\dagger a_k$) does not count particles but rather directly counts energy quantas. From there, the usual Fock quantization is known. It is useful however, for the sake of completeness, to develop it in a language closer to our originally found algebra, that is with:
\begin{equation}
\{\delta(k^2 + m^2)a_k, \delta(k'^2 + m^2)a_{k'}\} = -\mathrm{i}\sgn(k_t)\delta(k'^2 + m^2)\delta(k+k').
\end{equation}
This will lead to a more covariant expression more suited to the quantum gravity problem.

We must start with the one particle Hilbert space $\mathcal{H}$. First let $\mathbb{H}$ be the two-sheet hyperboloid embedded in $\mathbb{R}^3$ defined by:
\begin{equation}
t^2 - x^2 - y^2 = m^2
\end{equation}
where $(t,x,y)$ are the coordinates in $\mathbb{R}^3$. Now, $\mathcal{H}$ will be the space of functions from $\mathbb{H}$ into $\mathbb{C}$ equipped with the following scalar product:
\begin{equation}
\langle \psi | \phi \rangle = \int \delta(k^2 + m^2)\overline{\psi}(k)\phi(k) \mathrm{d}^3 k.
\end{equation}
This is the momentum representation for our one-particle. Because, we are interested in real valued fields, we will add the following constraint:
\begin{equation}
\forall k \in \mathbb{R}^3,\ \forall \phi \in \mathcal{H},\ \overline{\phi(k)} = -\phi(-k).
\end{equation}
Note the minus sign corresponding to the fact that $\overline{a_k} = -a_{-k}$. With this definition $\mathcal{H}$ is trivially a pre-Hilbertian space. By choosing a plane in $\mathbb{R}^3$ to parametrize $\mathbb{H}$, we get however that:
\begin{equation}
\langle \psi | \phi \rangle = \int \frac{1}{2\sqrt{\vec{k}^2 + m^2}}\overline{\psi}(k)\phi(k) \mathrm{d}^2 k.
\end{equation}
This shows that $\mathcal{H}$ is isomorphic to $\mathrm{L}^2(\mathbb{R}^2)$ with the caveat that the wave-functions must be divided $\sqrt{2E}$ in the mapping. This factor is actually quite important as it appeared in our algebra for the $a_k$ and this will allow a simpler representation of the creation-annihilation operators.

Now, we define the following sequence of Hilbert spaces:
\begin{enumerate}
	\item $\mathcal{H}_0 = \mathbb{C}$, the $0$-particle Hilbert space, also called the vacuum Hilbert space,
	\item $\mathcal{H}_1 = \mathcal{H}$, the $1$-particle Hilbert space as previously explained.
	\item $\mathcal{H}_n = \mathrm{Sym}(\mathcal{H}^{\otimes n})$, for $n \ge 2$, the symmetric part of the tensor product of $n$ copies of $\mathcal{H}$ and represents the $n$-particle Hilbert space for bosonic particles..
\end{enumerate}
The Fock space $\mathcal{H}_\phi$ is defined by:
\begin{equation}
\mathcal{H}_\phi = \bigoplus_{n\in\mathbb{N}} \mathcal{H}_n .
\end{equation}

Now, we can define the creation and annihilation operators $a_k$. There are two cases. First, let's consider $k$ such that $k^2 + m^2 = 0$ and $k_t < 0$. We define $\hat{a}_k$ by its restriction $\hat{a}_{k,n}$ on $\mathcal{H}_n$. For $n \ge 1$, we define $\hat{b}_{k,n}$:
\begin{equation}
\hat{b}_{k,n} : \left\{
\begin{array}{rcl}
\mathcal{H}^{\otimes n} &\rightarrow& \mathcal{H}^{\otimes (n-1)} \\
| v_1 \rangle \otimes | v_2 \rangle \otimes \cdots \otimes | v_n \rangle &\mapsto& \frac{1}{\sqrt{n}}\sum_{i=1}^n v_i(k) | v_1 \rangle \otimes | v_2 \rangle \otimes \cdots \otimes \widehat{| v_i \rangle} \otimes \cdots \otimes | v_n \rangle
\end{array}
\right.
\end{equation}
As standard, $\widehat{| v_i \rangle}$ means that $| v_i \rangle$ is omitted from the list. $\hat{a}_{k,n}$ is the restriction of $\hat{b}_{k,n}$ to $\mathcal{H}_n$. For $n=0$, we have:
\begin{equation}
\hat{a}_{k,0} : \left\{
\begin{array}{rcl}
\mathcal{H}_0 &\rightarrow& \mathcal{H}_{0} \\
v &\mapsto& 0
\end{array}
\right.
\end{equation}
which corresponds to the fact that the vacuum is annihilated by all annihilation operators.

Similarly, we can define $a_k$ for $k$ such as $k^2 + m^2 = 0$ and $k_t > 0$. This will act in the (algebraic) dual spaces. Let's define $\hat{b}_{k,n}$:
\begin{equation}
\hat{b}_{k,n} : \left\{
\begin{array}{rcl}
(\mathcal{H}^\star)^{\otimes n} &\rightarrow& (\mathcal{H}^\star)^{\otimes (n+1)} \\
\langle v_1 | \otimes \langle v_2 | \otimes \cdots \otimes \langle v_n | &\mapsto& \frac{1}{\sqrt{n+1}}\sum_{i=1}^{n+1} \langle v_1 | \otimes \langle v_2 | \otimes \cdots \otimes \langle u | \otimes \langle v_i | \otimes \cdots \otimes \langle v_n | ,
\end{array}
\right.
\end{equation}
with:
\begin{equation}
\forall | v \rangle \in \mathcal{H},\ \langle u | v \rangle = v(k).
\end{equation}
$\hat{a}_{k,n}$ is the restriction of $\hat{b}_{k,n}$ to $\mathcal{H}_n$. This concludes the matter sector.

\medskip

For the gravity sector, we have two sets of observables. We have the curvature constraints which, as long as we don't restrict to the constraint surface, are legitimate observables. We will write $D_I(x)$ from now on and remember that they are densities. And we have the triad $e_a^I(x)$. They are not exactly conjugate. The conjugate arise when we integrate them along a line (possibily starting from infinity as mentioned in section \ref{sec:basic_ops}). Then $\int^{P(\sigma)} e^I$ is conjugate to $D_I(x)$ and commutes with the $a$ operators. When we integrate, we loose some information. But it is remarkable that we don't loose gauge-invariant information: thanks to gauge-invariance, the integral of $e$ only depends on the end-point of the integral. That means we completely characterize the subspace defined by $de^I = 0$. This is this subspace that we will quantize.

The curvature constraints $D_I(x)$ are densities while, the integral of the triad acts as a scalar function. This setup is similar to Loop Quantum Gravity where conjugate quantities are carried by dual geometrical constructs. It is in fact exactly equivalent to the usual Loop Quantum Gravity setup except that here, because we have used gauge-invariant quantities, the support is on surfaces and points rather than lines. As a first approach however, we will not quantize in the standard fashion - that is using the Ashtekar-Lewandowski representation or its equivalent -  but will rather consider the equivalent of the BF representation \cite{Dittrich:2014wpa,Bahr:2015bra}. Indeed, we have two choices: either we start from a vacuum state where $e=0$ everywhere or we start with a vacuum state that has $D_I(x) = 0$ everywhere. The second case is akin to the BF vacuum and is very relevant to our problem: this vacuum state is precisely the solution to the constraints. So let's quickly sum up the construction in the abelian case.

Let's define the Hilbert space $\mathcal{H}_G$. Let $\mathcal{R}$ be the space of functions over $\Sigma$ valued in $\mathbb{R}^3$ that are zero everywhere except for a finite number of points. Now $\mathcal{H}_G$ is the space of square integrable functions over $\mathcal{R}$ equipped with the following scalar product:
\begin{equation}
\langle \Psi_1 | \Psi_2 \rangle = \sum_{\vec{f} \in \mathcal{R}} \overline{\Psi_1(\vec{f})} \Psi_2(\vec{f}).
\end{equation}
The sum is well-defined (though possibly infinite) thanks to the square integrable condition. Note that this space can be constructed by a projective limit (as it is standard in Loop Quantum Gravity). In that case, we would have functions depending on $\mathbb{R}^3$ labels for a finite number of points. Two functions with support on a different set of points would be equivalent (regarding cylindrical consistency) if they do not depend on the labels of the points that are no shared and if the dependency is the same for shared points. This is however not needed here thanks to the combination of two properties. First, because we look at the gauge-invariant subspace, the support is points rather than graph, things are greatly simplified. And because the gauge group is abelian, much simpler expressions can be given still. Nonetheless, the construction is similar in spirit: we have a normalized vacuum state which is:
\begin{equation}
\Psi_0(f) = \left\{
\begin{array}{rl}
1 &\textrm{if }f = \vec{0}, \\
0 &\textrm{otherwise.}
\end{array}
\right.
\end{equation}
Here $\vec{0}$ is understood to be the function that is constant over $\Sigma$ and equal to the vector $\vec{0}$. Then, excitations can be constructed with the action of the exponential of the integrated triad (which we will construct shortly). The Hilbert space is then the completion of the linear span of these excitations. This means that we have an Hilbertian basis given by the indicator functions once more. A member $\Psi_f$ of the basis is given for each function $f$ of $\mathcal{R}$ and is defined by:
\begin{equation}
\Psi_f(g) = \left\{
\begin{array}{rcl}
1 &\textrm{if}& g = f, \\
0 &\textrm{if}& g \neq f.
\end{array}
\right.
\end{equation}

The operator corresponding to $D_I(x)$ must be regularized. As $D_I(x)$ is a density, it is natural to consider the following integrated quantities: $\int N(x) D_I(x) \mathrm{d}^2\sigma$ where $N$ is some test function. We will therefore define the operator $\hat{D}_I[N]$. It is defined by its action on the basis in the following manner:
\begin{equation}
\hat{D}_I[N]\Psi_f = \left(\sum_{P \in \Sigma} N(P) f(P)_I\right)\Psi_f.
\end{equation}
This action is not always well-defined but it is on a dense subset of the space (namely the span of states $\Psi_f$ with functions $f$ that have finitely many non-zero points). We see here that the basis we constructed diagonalizes the $\hat{D}_I[N]$ operator. Similarly, we can defined the exponentiated operator for the triad. We do not need to regularize this time (except through the integral). Let $\vec{k}$ be in $\mathbb{R}^3$ and $P$ on $\Sigma$. We define $\hat{E}(\vec{k}, P)$ by its action of the basis:
\begin{equation}
\hat{E}(\vec{k}, P)\Psi_f = \Psi_{\tilde{f}},
\end{equation}
where:
\begin{equation}
\tilde{f}(Q) = \left\{
\begin{array}{rcl}
f(Q) &\textrm{if}& Q \neq P, \\
f(P)+\vec{k} &\textrm{if}& Q = P.
\end{array}
\right.
\end{equation}
As such $\hat{E}(\vec{k}, P)$ is the quantization of $\exp \left(-\mathrm{i}\vec{k}\cdot \int^P \vec{e}\right)$.

Note that the non-exponentiated version of the operator does not exist. In practice, this means we have used the Bohr compactification of $\mathbb{R}^3$ for the values of the integrals. This can be seen by the fact that the dual (present in eigenvalues of the curvature constraints) is $\mathbb{R}^3$ equipped with a discrete topology. This trick is handy to circumvent the problem of using non-compact groups. Sadly, the Bohr compactification is only injective for maximally almost periodic groups which the gauge group of the non-abelian theory ($\mathrm{SU}(1,1)$) is not. This is what prevents the standard Ashtekar-Lewandowski construction for non-compact gauge group. It should be noted however that such an obstruction is not present for the BF vacuum \cite{Bahr:2015bra}. It might very well be then, that the current construction generalizes to the non-abelian case.

Finally, the kinematical Hilbert space is simply $\mathcal{H}_G \otimes \mathcal{H}_\phi$ with the operators naturally extended. The solution to the constraints is simply: $(\mathbb{C} \Psi_0)\otimes \mathcal{H}_\phi \simeq \mathcal{H}_\phi$ where $\Psi_0$ is the vacuum for $\mathcal{H}_G$. It is trivial to see that this space is isomorphic to the standard Hilbert space for a free field theory. Though this construction is interesting to get a feel of how the theory works in the quantum realm, it is not satisfying on at least two accounts:
\begin{enumerate}
	\item First, it relies too much on a change of variable. Normally, to get a direct link with the classical theory, one would start with canonical variables and represent them, and then try to express constraints and similar operators. Here, not only have we not done that, it is not even possible to express the original operators. For instance, it is incredibly difficult (if not outright impossible) to extract the curvature operator out of the constraints. Indeed, to do that, we require both the fields operators (which we don't have) and the inverse of the metric (which does not even exist as an operator). Similarly, the natural expression for the momentum operator for the field depends on the normal operator, which does not exist because of the Bohr compactification we used.
	\item Second, it relies heavily on the abelian structure of the theory. All this approach was only possible because we can decouple completely two sectors that we might want to call the gravitational and the matter sector (though the curvature cosntraint has a bit of matter in it). This is not something we can hope for in a non-abelian theory. So the method is way too specific to our case.
\end{enumerate}
It does not mean it is not useful though: this acts as a guideline. We now know what the theory looks like and what to expect from different constructions.

The ideal construction however would start from the curvature operator, the triad and the field operators and then get the constraints. At least, it should be possible to reconstruct all these operators. This is however not possible in our case. Indeed, the curvature operator (or the holonomy operator) appears only in the curvature operator for now. As a consequence, we will first need the scalar field operator and the momentum operator to be able to retrieve it. However, from the work done in section \ref{subsec:Fourier}, we can use the Fourier transform in $\mathbb{R}^3$ to get expressions of $\phi$ and $\Pi$ in terms of the creation and annihilation operators. We get:
\begin{equation}
\left\{
\begin{array}{rcl}
\phi(\sigma) &=& \int \delta(k^2 + m^2) \frac{\sgn(k_t)}{\sqrt{2}\pi}a_k \mathrm{e}^{\mathrm{i} \vec{k} \cdot \int^\sigma \vec{e}} \mathrm{d}^3 k, \\
\Pi(\sigma) &=& \int \delta(k^2 + m^2) (\vec{k} \cdot \vec{n}) \frac{\sgn(k_t)}{\sqrt{2}\pi}a_k \mathrm{e}^{\mathrm{i} \vec{k} \cdot \int^\sigma \vec{e}} \mathrm{d}^3 k.
\end{array}
\right.
\end{equation}
The expression of $\Pi$ is particularly problematic as it relies on the existence of an operator for the normal $n$, which does not exists in our representation.

One might want to try and use the more standard Ashtekar-Lewandowski representation $\mathcal{H}_{AL}$. In that case, it is possible to construct a normal operator $n$ in a way similar to the area operator in LQG \cite{Ashtekar:1996eg}. However, in that case, we face another problem: given a state of the form $|0\rangle \otimes |\phi\rangle \in \mathcal{H}_{AL}\otimes\mathcal{H}_\phi$ where $|0\rangle$ is the AL vacuum and $|\phi\rangle$ is some state in $\mathcal{H}_\phi$, we have $\hat{\Pi}|0\rangle \otimes |\phi\rangle = 0$ irrespective of the state $|\phi\rangle$. This might be possible to cure, by forgetting about classical expressions and rather concentrating on reproducing the algebra at the quantum level. This would be however surprising since the expression for $\Pi$ is quite regular involving only exponentials and polynomials in the triad that commute among themselves and should not require regularization.

We want to suggest another direction in this paper, that we will start exploring in the next section. Though, we do not have a complete proof for a successful construction, the arguments we just laid out fail in this context. This solution, though it seems unnatural at first, has - in hindsight - geometrical justification. The idea is to use the work done by Koslowski and Sahlmann \cite{Koslowski:2007kh,Sahlmann:2010hn,Koslowski:2011vn} and to develop a representation peaked on a classical non-degenerate spatial metric. Though perfect diffeomorphism invariance (for the vacuum) is lost, there is still a notion of diffeomorphism covariance available and the geometrical interpretation we will offer justifies the choice of a particular background, at least for abelian gravity. We develop this approach in the following section.

\subsection{Ashtekar-Lewandowski representation peaked on a classical vacuum}

\label{sec:newrep}

The difficulty we face is linked to the non-existence of non-exponentiated versions of the triad operators on the Hilbert space. This is quite standard in Loop Quantum Gravity: the standard constructions only allow for one operator out of a conjugated pair to be defined, the other one is only defined through its exponentials. In the usual Ashtekar-Lewandowski representation \cite{Ashtekar:1996eg,Ashtekar:1997fb,Ashtekar:1998ak} for instance, the holonomy operators are well-defined but only the exponentiated versions are defined. In the BF representation defined by Dittrich \textit{et al.} \cite{Dittrich:2014wpa,Bahr:2015bra}, the triad is only defined through its exponentials, but some version of the logarithm of the holonomies are defined\footnote{There are in fact technical difficulties in this case because of the non-abelian nature of the gauge group. However, the limit for loops going to zero is usually well-defined (though group-valued) and play the same role.}. In our case, we have developed the equivalent of the BF representation, since the conjugate to the triad is defined. Moving to the standard Ashtekar-Lewandowski representation will not help however. Indeed, our problem is not only linked with the possibility of writing a simple triad operator but also the possibility of inverting it, at least to some extent as we want to be able to write the inverse determinant of the spatial metric. And the usual Ashtekar-Lewandowski representation does not allow for that (at least not in any known ways\footnote{Though Thiemann developed some ideas in this regard \cite{Thiemann:1996ay}, there are severe questions on whether his approach is successful \cite{Livine:2013wmq}.}) since the vacuum is degenerate everywhere and all the excited states are degenerate almost everywhere. If we want to write the inverse determinant, we will therefore need a new representation of the holonomy-flux algebra (or of its equivalent in our case - since we considered only the gauge-invariant sector).

It is noteworthy that some other representations have been discussed already in Loop Quantum Gravity, most notably \cite{Koslowski:2007kh,Sahlmann:2010hn,Koslowski:2011vn}. This representation is very similar to the Ashtekar-Lewandowski representation, except the vacuum is not peaked on degenerate geometry but rather on a given classical metric. Of course, diffeomorphism invariance of the vacuum is lost, which explains how the LOST theorem \cite{Lewandowski:2005jk} is evaded, and is replaced by a notion of diffeomorphism covariance. This representation is however very interesting to us because the metric is everywhere non-degenerate for the vacuum. Even for most of the excited states, the metric is non-degenerate and when it is not, it is only degenerate on a finite number of points. As long as we can reproduce the classical algebras correctly, this leads to very natural expressions for the inverse determinant of the metric. However, we have now traded another issue which is the choice of the background metric, which seems a bit counter-productive with regard to the standard Loop Quantum Gravity approach.

\medskip

Before tackling this problem however, let's sum up Koslowski's and Sahlmann's approach in \cite{Koslowski:2007kh,Sahlmann:2010hn,Koslowski:2011vn} and adapt it to our case. The construction uses the dual structure to the one we have done in section \ref{sec:firstapproach}. In the previous construction, the operators acting on surfaces (the constraints) were diagonal, and excitations were created by acting on points. Here, it is the reverse: the point operators are diagonal and the surface operators create the excitations. This means we need some projective techniques to deal with it correctly.

We can define a Hilbert space $\mathcal{H}_\Delta$ for a given triangulation $\Delta$ of $\Sigma$. This Hilbert space is the completion of the span of the basis given by $\mathbb{R}^3$ labels of the triangles that are non-zero for a only finite number of triangles. We can make this precise in the following manner: let $\mathcal{F}_\Delta$ be the space of functions for the triangles of $\Delta$ into $\mathbb{R}^3$ such that the values are non-zero for a finite number of triangles. This is the space of labels on the triangulation. The elements of $\mathcal{H}_\Delta$ are functions from $\mathcal{F}_\Delta$ into $\mathbb{C}$ that are square integrable for:
\begin{equation}
\langle \psi | \phi \rangle = \sum_{f \in \mathcal{F}_\Delta} \overline{\psi(f)} \phi(f).
\end{equation}
The full (continuous) Hilbert space is defined as:
\begin{equation}
\mathcal{H}_{KS} = \left(\bigcup_{\Delta} \mathcal{H}_\Delta \right)\Big\slash \sim.
\end{equation}
Here the union is a disjoint union over all possible triangulations of $\Sigma$. We must now define the equivalence relation~$\sim$.

For this, we need the notion of a refinement of a triangulation. A triangulation $\Delta'$ is a refinement of $\Delta$ if for any triangle in $\Delta$ is the union of triangles in $\Delta'$. We can then map any function of $\mathcal{F}_\Delta$ into $\mathcal{F}_{\Delta'}$. For $f \in \mathcal{F}_\Delta$, we define $f' \in \mathcal{F}_{\Delta'}$ as:
\begin{equation}
f'(t) = f(T)\textrm{, with }t \subseteq T.
\end{equation}
Similarly, we can write extend a state $\psi \in \mathcal{H}_\Delta$ into $\psi' \in \mathcal{H}_{\Delta'}$ as follows:
\begin{equation}
\psi'(f) = \left\{
\begin{array}{rl}
\psi(g)&\textrm{if }g'=f, \\
0 &\textrm{otherwise.}
\end{array}\right.
\end{equation}
We can finally get to our equivalence relation necessary to define $\mathcal{H}_{KS}$. Two states $\psi \in \mathcal{h}_\Delta$ and $\psi' \in \mathcal{H}_{\Delta'}$ are equivalent if and only if there exists a refinement $\Delta''$ of both $\Delta$ and $\Delta'$ such that the extension of $\psi$ and $\psi'$ in $\mathcal{H}_{\Delta''}$ are identical. Note that if this is true, it is true for any refinement of both triangulations. Note also that there is always a refinement of both triangulations but there is no guarantee that the extension of $\psi$ and $\psi'$ will match.

Up to this point, the definition actually follows the techniques of the BF vacuum in order to adapt the construction to quantities carried by surfaces and points (rather than lines). But what will distinguish $\mathcal{H}_{KS}$ from both the BF representation and the standard AL representation is the construction of the operators.

First, let's start with the simplest operator: the integrated curvature constraint. Let $\Delta$ be a triangulation of $\Sigma$ and $\phi$ a function from the triangles into $\mathbb{R}^3$ non-zero only a finite number of triangles. If $\Delta'$ is a refinement of $\Delta$, we define:
\begin{equation}
\widehat{\mathrm{e}^{\mathrm{i} D[\phi]}} : \left\{
\begin{array}{rcl}
\mathcal{H}_{\Delta'} &\rightarrow& \mathcal{H}_{\Delta'} \\
\psi &\mapsto& \psi'
\end{array}
\right.
\end{equation}
with:
\begin{equation}
\psi'(f) = \psi(f + \phi).
\end{equation}
The final sum is done by extending $\phi$ to $\Delta'$. This is standard action, completely equivalent, so far, to the one in the AL-representation. This action can be extended on coarser representation. It is compatible with the quotient and therefore carries to whole space $\mathcal{H}_{KS}$.

Second, we can consider the triad operator. This is done in two steps. As a first step, let $\Delta$ be a triangulation. Let's denote$| \psi_f \rangle$ the state in $\mathcal{H}_{\Delta}$ defined by:
\begin{equation}
\psi_f(g) = \left\{
\begin{array}{rl}
1 &\textrm{if }f=g,\\
0 &\textrm{otherwise,}
\end{array}
\right.
\end{equation}
with $f \in \mathcal{F}_{\Delta}$. These states form a (Hilbertian) basis of $\mathcal{H}_{\Delta}$. We can now define:
\begin{equation}
\widehat{\mathcal{O}[\phi]} | \psi_f \rangle = \sum_{\sigma \in \Sigma} \phi(\sigma) \cdot f(\sigma) | \psi_f \rangle,
\end{equation}
with $\phi$ is a function from $\Sigma$ into $\mathbb{R}^3$ with finitely many non-zero values. Thus $\sum_{\sigma \in \Sigma} \phi(\sigma) \cdot f(\sigma)$ is understood as a sum over these finitely many values and $f(\sigma)$ is the label for the triangle of $\Delta$ that $\sigma$ belongs to\footnote{In practice, this means that this sums is not well-defined if the point $\sigma$ fulls on an edge or a vertex of the triangulation. This is not important for us as we can just reduce the domain of the operator.}. We recognize here the definition of the triad operator in the standard AL-representation. But now, as a second step, let's define a background field $\tilde{e} : \Sigma \rightarrow \mathbb{R}^3$. And consider the following operator:
\begin{equation}
\widehat{e[\phi]} = \int \phi\cdot\tilde{e} + \widehat{\mathcal{O}[\phi]}.
\end{equation}
This operator trivially has the same algebra but is peaked on a classical configuration for the triad. This is the main difference of the KS representation (compared to the usual AL one).

\medskip

Now, all this construction relies on a choice of background metric and even, to be more precise, a choice of background triad. This choice seems arbitrary at first, but in our case there is a very natural way to select a class of metrics. Indeed, we have to remember that we are considering the gauge-invariant subspace which translates to the condition:
\begin{equation}
\mathrm{d}\mathrm{e}^I = 0.
\end{equation}
This condition entails that, if we restrict once more to a simply connected manifold, the triad derives from a potential $\Psi^I$. This functions acts as an embedding of $\Sigma$ into $\mathbb{R}^3$ (if the metric is invertible). But it also means that the integrated triad is zero on any closed loops. And this is valid also on the vacuum state. This means that the background triad must satisfy all these conditions and in particular correspond to an embedding into $\mathbb{R}^3$. Up to topological questions, that we have discarded as we are considering the simplest case, this means that the metric is fixed up to diffeomorphism. This entails in turn that the construction will indeed depend on the metric but once the diffeomorphism constraints will be enforced, diffeomorphism invariance will be restored in a way which is independent from the choice of the initial metric (as long as it is invertible). So, from now on, let's just choose a background embedding into $\mathbb{R}^3$ and use the triad that derives from it.

Let's turn back to the full representation, including the matter sector. Our goal was to able to write expressions like:
\begin{equation}
\left\{
\begin{array}{rcl}
\phi(\sigma) &=& \int \delta(k^2 + m^2) \frac{\sgn(k_t)}{\sqrt{2}\pi}a_k \mathrm{e}^{\mathrm{i} \vec{k} \cdot \int^\sigma \vec{e}} \mathrm{d}^3 k, \\
\Pi(\sigma) &=& \int \delta(k^2 + m^2) (\vec{k} \cdot \vec{n}) \frac{\sgn(k_t)}{\sqrt{2}\pi}a_k \mathrm{e}^{\mathrm{i} \vec{k} \cdot \int^\sigma \vec{e}} \mathrm{d}^3 k.
\end{array}
\right.
\end{equation}
This suggested that the gravity sector needed a new representation. The Fock space used for matter is however completely equipped for such expressions. We will therefore rather keep it. This leads to the full Hilbert space:
\begin{equation}
\mathcal{H}_{\textrm{Full}} = \mathcal{H}_{KS}\otimes\mathcal{H}_\phi.
\end{equation}
Before moving to the next section, let's make a final remark: though this representation gives natural inverse operators, in a sense, this does not matter. What matters is the algebra of the operators. In the end, we must find two natural pairs of collections of operators, corresponding to the field and momentum operator on the one hand and to the triad and curvature operator on the other. Moreover, these operator should lead to expressions for the constraints that match the previously found algebra. If the naive inversion fails, this will mean that this technique fails. This is what in the end should guide such construction. And these tests are still to be done with the method we just suggested.

\section{Discussion \& Future work}

Granted the previous idea can be made to work, the natural question is whether this can be extended outside of the abelian theory. Indeed, the representation we chose depended on a background which, for the abelian case, can be chosen naturally. This however depended on the resolution of the Gauß constraints. In the non-abelian case, such a procedure might not be that well-defined. A few points are encouraging though: this representation gives a natural understanding of how matter propagates on an (abelian) quantum spacetime. Indeed, as we mentioned early on in this paper, the theory we developed is, at least in some sector, equivalent to a free scalar field theory. With such a theory, spacetime is completely classical. Our theory however is completely quantum mechanical, including spacetime. On the constraint surface, the triad in particular is completely ill-defined (in a quantum mechanical sense) and only the curvature has a precise value. We might wonder how a field might propagate freely here. The answer, according to the construction we have just done, is simple: spacetime really is flat. The degeneracy of the triad does not come from a true quantum degeneracy but rather is caused by the superposition of all the states coming from the action of the diffeomorphism constraints. The final state therefore is a superposition of classical flat space but seen from all possible coordinate systems. This is of course possible only because there are no local degrees of freedom in 3d gravity. Though, it might be possible to extend these techniques to non-abelian 3d gravity, the implications are not quite as clear for the 4d case. An interesting idea, that has been explored almost accidentally in the context of cosmology (see for instance \cite{mukhanov2007introduction}) as a first approach, is that only local degrees of freedom (that is gravitational waves) are quantum in that sense.

\medskip

Let's get back to the 3d problem. Even in that case, once we want to get to the full non-abelian theory, a few roadblocks appear. One of the major problem is path-dependency. Indeed, we defined the following operator as a creation operator:
\begin{equation}
a_k = \frac{1}{\sqrt{2}\pi}\int \left(k^I n_I \phi + \mathrm{i}\Pi\right) \mathrm{e}^{- \mathrm{i} \vec{k} \cdot \int^{\sigma} \vec{e}} \mathrm{d}^2\sigma .
\end{equation}
There we used the integral $\int^\sigma \vec{e}$ which did not depend on the path chosen as long as the Gauß constraints were satisfied. A natural extension to the non-abelian case would be:
\begin{equation}
b_k = \frac{1}{\sqrt{2}\pi}\int \left(k^I n_I \phi + \mathrm{i}\Pi\right) \mathrm{e}^{- \mathrm{i} \vec{k} \cdot \int^{\sigma} g \triangleright \vec{e}} \mathrm{d}^2\sigma ,
\end{equation}
where $g$ is the holonomy of the connection along the integration path and acts as parallel transport. Though this expression is gauge-invariant, it depends on the path chosen for the integration, even when the Gauß constraints are checked. This makes the correct generalization quite unclear. Two points should be underlined here however. First, similar problem have been dealt with in the construction of the BF representation and have been solved by a systematic choice of paths for gauge-fixing \cite{Dittrich:2014wda}. This is moreover close to book-keeping techniques needed for the classical solution of the problem \cite{tHooft:1992izc} which seems to support such an approach. Second, this problem can be partially recovered in the abelian case, if one wants to define the theory more generally without imposing first the Gauß constraints. This might be needed anyway to be able to check the brackets of all the quantum operators we are interested in from the end of section \ref{sec:newrep}. This will therefore be an interesting intermediate step to consider.

\medskip

The abelian case also relied on the commutativity between the operators $a_k$ and the constraints $D_I$. It would be surprising, to say the least, that such a setup could be possible in the non-abelian case, for the operators $b_k$ and the corresponding constraints $\tilde{D}_I$. Several scenarios can be envisioned, the most probable to our eyes though is that, though the $b_k$ will not commute with the constraints, it should still be possible to make them into the algebra of creation and annihilation operator for some non-commutative field theory. In that case, they would allow us to write a basis of states on which it is reasonable to to a perturbative study. Ideally of course, some exact cases could be found, like a $m \rightarrow 0$ limit, one-particle states or maybe some cosmological setup. In any case, the non-commutativity is not a problem as long as we can interpret it to be almost commutative in some limit. This, however, will only be possible if we can develop the full set of operators $\phi$, $\Pi$, $e$ and $A$ independently from the techniques we have employed in the commutative case. This means that one of the most important point moving forward is concluding the program opened by section \ref{sec:newrep}.

\medskip

Let's mention one last point before wrapping up: the idea of studying the abelian theory as a starting point, possibly for perturbative expansion is not new and was originally introduced by Smolin \cite{Smolin:1992wj}. In our case however, we wanted it in particular to be able to study the geometry of the quantum spacetime. According to Connes'work (for instance \cite{Chamseddine:1996zu}), this is better encoded in the Dirac operator governing the propagation of fermions rather than just the metric. A similar approach would then start with fermions coupled to abelian gravity. This is however rather ill-defined at the moment. Indeed, the gauge group does share the same topology as $\mathrm{SU}(1,1)$, making the distinctions between bosons and fermions less clear. Moreover, it is not completely straightforward how the abelian connection should be coupled to the fermions. This is therefore an interesting point to explore further in future work.

\section{Conclusion}

In this paper, we considered a simplified model for 3d quantum gravity coupled to a scalar field. The model was taken from Smolin work \cite{Smolin:1992wj}, corresponds to a specific $G \rightarrow 0$ limit of standard 3d gravity, and can be formulated as standard BF theory (coupled to a scalar field in our case) but with an abelian gauge group. In four dimensions, this corresponds to a linearization of gravity but still expressed in a diffeomorphism invariant way. In three dimensions, the theory is still topological, but the dynamics is simplified. We showed in this paper in particular that a full sector of the theory is completely equivalent to a free scalar field, the gravity field only being there to allow for a diffeomorphism covariant formulation. This sector is actually fairly similar to what was already developed with parametrized field theories \cite{Kuchar:1989bk,Kuchar:1989wz,Varadarajan:2006am}, although in higher dimensions and with a different language.

\medskip

We showed furthermore that this equivalence with a free scalar field theory leads to the formulation of a creation-annihilation algebra of operators, even in a diffeomorphism invariant setting. This algebra can in principle be extended to other sectors of the theory as long as the metric is everywhere invertible. Though the natural formulation is a bit different due to diffeomorphism invariance, the algebra is completely equivalent to the standard one for the free scalar field. The interesting point is that all these operators commute with the constraints for the abelian theory. This means they allow the construction of a set of solutions of the constraints, and mirror the fact that the classical abelian gravity theory (coupled to a scalar field) is equivalent, at least in some sector, to the classical free scalar field theory. This also means that these expressions are a good starting point for studying the non-abelian theory, for instance to try and quantize the theory perturbatively. This also allows the construction of a full quantization of the linear theory based on these operators as new variables. The quantum theory splits into two sectors. One is the sector that encodes the various excitation of the scalar field, and can be mapped one to one to the free scalar field theory. The second can be understood as the gravity sectors that more or less decouples in this abelian theory. It can be mapped onto the BF theory and be solved exactly.

\medskip

The drawback of such an approach is that some natural operators do not exist or are extremely difficult to construct. In particular, the momentum operator for the scalar field, and the holonomy operator for the gravity field, require the definition of (non-exponentiated) triad operators and an inverse-metric operator. This implies in particular, that even the canonical variables of the theory cannot be expressed simply or may be downright impossible to write. This is not really a specific problem of our approach: we used the equivalent of the BF representation in our construction which has similar difficulties for constructing triad operators or inverse-triad operators. However, in our case, these difficulties become a problem when trying to write a correlation operator for the scalar field for instance, which is a quantity we will eventually want to be able to compute. Using the older and somewhat more standard Ashtekar-Lewandowski representation only partially solves the problem. If it is indeed possible to define a non-exponentiated triad operator, the fact that the metric is degenerate almost everywhere for almost all states create huge problems with our approach which precisely requires the opposite. Moreover, natural expressions for the momentum operator of the scalar field are pathological, even though they only require exponential and polynomial terms in the triads, which should not need any regularization for the quantum case.

\medskip

We offered a possible way out. Though the construction needs to be studied more thoroughly, the drawbacks of the previous two approaches disappear. The idea is to construct a representation peaked on a given classical state for the spatial metric. This idea was explored by Koslowski and Sahlmann \cite{Koslowski:2007kh,Sahlmann:2010hn,Koslowski:2011vn} as an equivalent to condensed state around a classical configuration. Though strict diffeomorphism invariance of the vacuum was lost, a sens of diffeomorphism covariance can still be retained. However, if this breaking was natural in their case, it seems more dubious when studying the theory from a more fundamental standpoint. We showed however that a specific vacuum can be selected using the Gauß constraints in the linear case and corresponds to a flat space. Because the vacuum is nowhere degenerate, all the problems with the previously mentioned representations are lifted. Interestingly, the construction also allows a very nice interpretation of how the spacetime on which a free scalar field propagates is recovered in a setup where the triad is supposed to be completely degenerate in a quantum sense. In fact (in the abelian case), the classical spacetime is there all along and the degeneracy only comes from the superposition of all the diffeomorphism equivalent way of describing the system.

\medskip

Finally, we left several questions open for further inquiry. Most notably, as we just said, the new representation we offered should be studied further. Indeed, even though the straightforward problems have been lifted, the study of the construction of the full operator set is still to be done. We left it ou however because a full and complete study would include a more complete treatment of the Gauß constraints which we just assume to be satisfied. Lifting this condition requires dealing with gauge fixing, choice of path when integrating, etc. These points must be considered at some point as they are needed for the non-abelian theory but were left out of this first investigation. Similarly, we have left out all questions regarding the various possible sectors of the theory, the role of topology, the possible restrictions when considering compact spaces, etc. Though this is certainly worth investigating on its own merit, our goal was to get a first grap on how to develop a non-abelian theory. In this regard, though all this is very important, it will most probably be quite different when changing the Lorentz gauge group.

\section*{Acknowledgements}

I would like to thank Stefan Hohenegger for the numerous discussions that helped and guided this project, and without whom none of this would have been possible. I would also like to thank John Barrett for the various conversations that launched the initial idea for this paper.

\appendix

\section{Details of the Hamiltonian analysis}
\label{app:hamil}

\subsection{Primary constraints and Hamiltonian}

We have the following action as a starting point:
\begin{eqnarray}
S[e,A,\phi] &=& \int_{\mathcal{S}} \Big[ \frac{\alpha}{2} \epsilon_{IJK} \epsilon^{\mu\nu\rho} e_\mu^I (\partial_\nu A_\rho^{JK} - \partial_\rho A_\nu^{JK}) + \frac{\Lambda}{6} \epsilon_{IJK} \epsilon^{\mu\nu\rho} e_\mu^I e_\nu^J e_\rho^K \nonumber \\
&-& \frac{1}{12} \epsilon_{IJK} \epsilon^{\mu\nu\rho} e_\mu^I e_\nu^J e_\rho^K \left( e^\sigma_M e^\tau_N \eta^{MN} \right) \partial_\sigma \phi \partial_\tau \phi - \frac{m^2}{12} \epsilon_{IJK} \epsilon^{\mu\nu\rho} e_\mu^I e_\nu^J e_\rho^K \phi^2 \Big] \mathrm{d}^3 x .
\end{eqnarray}
Let's start the hamiltonian analysis by choosing a integration manifold. We will simply choose $\mathcal{S} = \mathbb{R}^3$ to avoid some problems on compact manifolds and with non-trivial topology. We will though also neglect boundary terms, assuming nice behaviour at infinity.

Let choose some decomposition of $\mathcal{S}$ as $\mathbb{R}\times\Sigma$ with corresponding coordinates $(t,\sigma)$. $t$ will be our time variable and $\sigma$ will be the coordinates on the spatial slice $\Sigma$. We only assume that $\Sigma$ is diffeomorphic to $\mathbb{R}^2$ but not that it is a flat slice.

This allows the following writing:
\begin{equation}
S[e,A,\phi] = \int_\mathbb{R} L \mathrm{d}t,
\end{equation}
with:
\begin{eqnarray}
L &=& \int_{\Sigma} \Big[ \frac{\alpha}{2} \epsilon_{IJK} \epsilon^{\mu\nu\rho} e_\mu^I (\partial_\nu A_\rho^{JK} - \partial_\rho A_\nu^{JK}) + \frac{\Lambda}{6} \epsilon_{IJK} \epsilon^{\mu\nu\rho} e_\mu^I e_\nu^J e_\rho^K \nonumber \\
&-& \frac{1}{12} \epsilon_{IJK} \epsilon^{\mu\nu\rho} e_\mu^I e_\nu^J e_\rho^K \left( e^\sigma_M e^\tau_N \eta^{MN} \right) \partial_\sigma \phi \partial_\tau \phi - \frac{m^2}{12} \epsilon_{IJK} \epsilon^{\mu\nu\rho} e_\mu^I e_\nu^J e_\rho^K \phi^2 \Big] \mathrm{d}^2 \sigma .
\end{eqnarray}
We can now define the various momenta.

Let's note $B$ the momentum conjugated to $A$, $X$ the momentum conjugated to $e$ and $\Pi$ the momentum conjugated to $\phi$. The definitions are:
\begin{equation}
\left\{
\begin{array}{rcl}
B^\mu_{IJ}(\sigma) &\equiv& \frac{\delta L}{\delta (\partial_0 A_\mu^{IJ}(\sigma))},\\
X^\mu_I(\sigma) &\equiv& \frac{\delta L}{\delta (\partial_0 e_\mu^I(\sigma))},\\
\Pi(\sigma) &\equiv& \frac{\delta L}{\delta (\partial_0 \phi(\sigma))}.
\end{array}
\right.
\end{equation}
Here, it is understood that $\partial_0$ means derivative with respect to the time variable $t$.

This leads to our primary constraints. Let's start with the easy ones:
\begin{equation}
X^\mu_I = 0.
\end{equation}
This comes from the fact that the action does not depend at all on the derviatives of $e$.

Let's now turn to the variable $B$. We must distinguish two cases. First, $B^0$ is easy to study as no time derivate of $A_0$ appears in the action. Therefore:
\begin{equation}
B^0_{IJ} = 0.
\end{equation}
The story is a bit different for $B^a$ ($a \neq 0$). Here we rather get:
\begin{equation}
B^a_{IJ} = 2 \alpha \epsilon_{KIJ} \epsilon^{\mu 0 a} e_\mu^K = 2\alpha \epsilon_{IJK} \epsilon^{ab} e_b^K.
\end{equation}

There is no constraint on $\Pi$ as the relation we get is invertible in $\partial_0 \phi$. More precisely, we get:
\begin{equation}
\Pi = -\frac{1}{6} \epsilon_{IJK} \epsilon^{\mu\nu\rho} e_\mu^I e_\nu^J e_\rho^K \left( e^0_M e^\tau_N \eta^{MN} \right) \partial_\tau \phi = -(\det e)g^{0\tau}\partial_\tau \phi .
\end{equation}
This can be inverted into:
\begin{equation}
\partial_0 \phi = -\frac{1}{(\det e)g^{00}}\left(\Pi + (\det e)g^{0a}\partial_a \phi\right) .
\end{equation}
We have assumed here that the metric is invertible.

We can, at last, write the Hamiltonian which is defined as:
\begin{equation}
H \equiv \int_\Sigma \left(\frac{1}{2} B^\mu_{IJ} \partial_0 A_\mu^{IJ} + X^\mu_I \partial_0 e_\mu^I + \Pi \partial_0 \phi \right) \mathrm{d}^2 \sigma - L.
\end{equation}
Thanks to the constraints, most of the first terms vanish. We will only get the $\Pi$ term, as well as the $B^a$ terms. At the end of the day, we must also make sure that the final expression does not depend on $\partial_0 \phi$. We must therefore take some time to rewrite $L$ so that any time component is made explicit and not bulked together with the spatial ones.

So let's try to declutter $L$ a bit:
\begin{eqnarray}
L &=& \int_{\Sigma} \Big[ \frac{\alpha}{2} \epsilon_{IJK} \epsilon^{\mu\nu\rho} e_\mu^I (\partial_\nu A_\rho^{JK} - \partial_\rho A_\nu^{JK}) + \frac{\Lambda}{6} \epsilon_{IJK} \epsilon^{\mu\nu\rho} e_\mu^I e_\nu^J e_\rho^K \nonumber \\
&-& \frac{1}{12} \epsilon_{IJK} \epsilon^{\mu\nu\rho} e_\mu^I e_\nu^J e_\rho^K \left( e^\sigma_M e^\tau_N \eta^{MN} \right) \partial_\sigma \phi \partial_\tau \phi - \frac{m^2}{12} \epsilon_{IJK} \epsilon^{\mu\nu\rho} e_\mu^I e_\nu^J e_\rho^K \phi^2 \Big] \mathrm{d}^2 \sigma \nonumber \\	
&=& \int_{\Sigma} \Big[ e_0^I \frac{\alpha}{2} \epsilon_{IJK} \epsilon^{ab} (\partial_a A_b^{JK} - \partial_b A_a^{JK}) + \alpha \epsilon_{IJK} \epsilon^{ab} e_b^I \partial_0 A_a^{JK} + \alpha \epsilon_{IJK} \epsilon^{ab} e_a^I \partial_b A_0^{JK} \nonumber \\
&+& e_0^I \frac{\Lambda}{2} \epsilon_{IJK} \epsilon^{ab} e_a^J e_b^K - e_0^I \frac{1}{4} \epsilon_{IJK} \epsilon^{ab} e_a^J e_b^K g^{00} \partial_0 \phi \partial_0 \phi \nonumber \\
&-& e_0^I \frac{1}{2} \epsilon_{IJK} \epsilon^{ab} e_a^J e_b^K \left( e^0_M  e^c_N \eta^{MN} \right) \partial_0 \phi \partial_c \phi \nonumber \\
&-& e_0^I \frac{1}{4} \epsilon_{IJK} \epsilon^{ab} e_a^J e_b^K \left( e^c_M  e^d_N \eta^{MN} \right) \partial_c \phi \partial_d \phi \nonumber \\
&-& e_0^I \frac{m^2}{4} \epsilon_{IJK} \epsilon^{ab} e_a^J e_b^K \phi^2 \Big] \mathrm{d}^2 \sigma .
\end{eqnarray}
Now assuming we can neglect the condition at the boundary (for example by asking all the fields to vanish at infinity), we can rewrite this a bit:
\begin{eqnarray}
L &=& \int_{\Sigma} \Big[ \frac{1}{2} \partial_0 A_a^{IJ} \left(2 \alpha \epsilon_{IJK} \epsilon^{ab} e_b^K \right) + \frac{1}{2} A_0^{JK} \left(- 2 \alpha \epsilon_{IJK} \epsilon^{ab} \partial_b e_a ^I \right)  \nonumber \\
&+& e_0^I \Big(\alpha \epsilon_{IJK} \epsilon^{ab} F_{ab}^{JK}[A] + \frac{\Lambda}{2} \epsilon_{IJK} \epsilon^{ab} e_a^J e_b^K - \frac{1}{4} \epsilon_{IJK} \epsilon^{ab} e_a^J e_b^K g^{00} \partial_0 \phi \partial_0 \phi \nonumber \\
&-& \frac{1}{2} \epsilon_{IJK} \epsilon^{ab} e_a^J e_b^K \left( e^0_M  e^c_N \eta^{MN} \right) \partial_0 \phi \partial_c \phi \nonumber \\
&-& \frac{1}{4} \epsilon_{IJK} \epsilon^{ab} e_a^J e_b^K \left( e^c_M  e^d_N \eta^{MN} \right) \partial_c \phi \partial_d \phi \nonumber \\
&-& \frac{m^2}{4} \epsilon_{IJK} \epsilon^{ab} e_a^J e_b^K \phi^2 \Big) \Big] \mathrm{d}^2 \sigma .
\end{eqnarray}
Let's now define :
\begin{equation}
n_I = \frac{1}{2} \epsilon_{IJK} \epsilon^{ab} e_a^J e_b^K .
\end{equation}
This will allow the following more compact expression:
\begin{eqnarray}
L &=& \int_{\Sigma} \Big[ \frac{1}{2} \partial_0 A_a^{IJ} \left(2 \alpha \epsilon_{IJK} \epsilon^{ab} e_b^K \right) + \frac{1}{2} A_0^{JK} \left(- 2 \alpha \epsilon_{IJK} \epsilon^{ab} \partial_b e_a ^I \right)  \nonumber \\
&+& e_0^I \Big(\alpha \epsilon_{IJK} \epsilon^{ab} F_{ab}^{JK}[A] + \Lambda n_I - \frac{1}{2} n_I g^{00} \partial_0 \phi \partial_0 \phi \nonumber \\
&-& n_I \left( e^0_M  e^c_N \eta^{MN} \right) \partial_0 \phi \partial_c \phi \nonumber \\
&-& \frac{1}{2} n_I \left( e^c_M  e^d_N \eta^{MN} \right) \partial_c \phi \partial_d \phi \nonumber \\
&-& \frac{m^2}{2} n_I \phi^2 \Big) \Big] \mathrm{d}^2 \sigma .
\end{eqnarray}
Let's go back to the Hamiltonian. We have:
\begin{eqnarray}
H &=& \int_{\Sigma} \Big[ \frac{1}{2} \partial_0 A_0^{IJ} B^0_{IJ} + \frac{1}{2} \partial_0 A_a^{IJ} \left( B^a_{IJ}  - 2 \alpha \epsilon_{IJK} \epsilon^{ab} e_b^K\right) \nonumber \\
&+& X^\mu_I \partial_0 e_\mu^I + \Pi \partial_0 \phi - \frac{1}{2} A_0^{JK} \left(- 2 \alpha \epsilon_{IJK} \epsilon^{ab} \partial_b e_a ^I \right)  \nonumber \\
&-& e_0^I \Big(\alpha \epsilon_{IJK} \epsilon^{ab} F_{ab}^{JK}[A] + \Lambda n_I - \frac{1}{2} n_I g^{00} \partial_0 \phi \partial_0 \phi \nonumber \\
&-& n_I \left( e^0_M  e^c_N \eta^{MN} \right) \partial_0 \phi \partial_c \phi - \frac{1}{2} n_I \left( e^c_M  e^d_N \eta^{MN} \right) \partial_c \phi \partial_d \phi \nonumber \\
&-& \frac{m^2}{2} n_I \phi^2 \Big) \Big] \mathrm{d}^2 \sigma .
\end{eqnarray}
In this expression, we must now write $\partial_0 \phi$ in terms of $\Pi$ using:
\begin{equation}
\partial_0 \phi = -\frac{1}{(\det e)g^{00}}\left(\Pi + (\det e)g^{0a}\partial_a \phi\right) .
\end{equation}
Lets concentrate only on the relevant terms $T$:
\begin{eqnarray}
T &\equiv& \Pi \partial_0 \phi + \frac{1}{2}  e_0^I n_I g^{00} \partial_0 \phi \partial_0 \phi + e_0^I n_I \left( e^0_M  e^c_N \eta^{MN} \right) \partial_0 \phi \partial_c \phi \nonumber \\
&=& -\frac{1}{(\det e)g^{00}}\Pi \left(\Pi + (\det e)g^{0a}\partial_a \phi\right) + \frac{1}{2} e_0^I n_I g^{00} \left[\frac{1}{(\det e)g^{00}}\left(\Pi + (\det e)g^{0a}\partial_a \phi\right)\right]^2 \nonumber \\
&-& e_0^I n_I \left( e^0_M  e^c_N \eta^{MN} \right)\frac{1}{(\det e)g^{00}}\left(\Pi + (\det e)g^{0a}\partial_a \phi\right)\partial_c \phi \nonumber \\
&=& -\frac{1}{2(\det e)g^{00}}\Pi^2 - \frac{1}{g^{00}} \left( e^0_M  e^c_N \eta^{MN} \right) \Pi \partial_c \phi \nonumber \\
&+& \frac{\det e}{2g^{00}} g^{0a} g^{0b} \partial_a \phi \partial_b \phi - \frac{\det e}{g^{00}} \left( e^0_M  e^c_N \eta^{MN} \right) g^{0a} \partial_a \phi \partial_c \phi \nonumber \\
&=& -\frac{1}{2(\det e)g^{00}}\Pi^2 - \frac{g^{0c}}{g^{00}} \Pi \partial_c \phi - \frac{\det e}{2g^{00}} \left(g^{0a} \partial_a \phi\right)^2 .
\end{eqnarray}
Let's put this in one single package:
\begin{eqnarray}
H &=& \int_{\Sigma} \Big[ \frac{1}{2} \partial_0 A_0^{IJ} B^0_{IJ} + \frac{1}{2} \partial_0 A_a^{IJ} \left( B^a_{IJ}  - 2 \alpha \epsilon_{IJK} \epsilon^{ab} e_b^K\right) + X^\mu_I \partial_0 e_\mu^I - \frac{1}{2} A_0^{JK} \left(- 2 \alpha \epsilon_{IJK} \epsilon^{ab} \partial_b e_a ^I \right)  \nonumber \\
&-& e_0^I \Big(\alpha \epsilon_{IJK} \epsilon^{ab} F_{ab}^{JK}[A] + \Lambda n_I - \frac{1}{2} n_I g^{cd} \partial_c \phi \partial_d \phi - \frac{m^2}{2} n_I \phi^2 + \frac{n_I}{2g^{00}} (g^{0a}\partial_a \phi)^2 \Big) \nonumber \\
&-& \frac{1}{2(\det e)g^{00}}\Pi^2 - \frac{g^{0c}}{g^{00}} \Pi \partial_c \phi \Big] \mathrm{d}^2 \sigma .
\end{eqnarray}
Note that:
\begin{equation}
g^{cd} - \frac{g^{0c} g^{0d}}{g^{00}} = h^{cd}
\end{equation}
where $h^{cd}$ denotes the inverse of the induced metric on $\Sigma$. In particular, it does not depend on $e_0^I$. Similarly:
\begin{equation}
\frac{1}{(\det e) g^{00}} = \frac{1}{(\det e) \frac{\det h}{\det g}} = -\frac{\det e}{\det h},
\end{equation}
which is linear in $e_0^I$. We can see therefore that every single one of the last terms is linear in $e_0^I$. We can sum up this in the following formula:
\begin{eqnarray}
H &=& \int_{\Sigma} \Big[ \frac{1}{2} \partial_0 A_0^{IJ} B^0_{IJ} + \frac{1}{2} \partial_0 A_a^{IJ} \left( B^a_{IJ}  - 2 \alpha \epsilon_{IJK} \epsilon^{ab} e_b^K\right) + X^\mu_I \partial_0 e_\mu^I - \frac{1}{2} A_0^{JK} \left(- 2 \alpha \epsilon_{IJK} \epsilon^{ab} \partial_b e_a ^I \right)  \nonumber \\
&-& e_0^I \Big(\alpha \epsilon_{IJK} \epsilon^{ab} F_{ab}^{JK}[A] + \Lambda n_I - \frac{1}{2} n_I h^{cd} \partial_c \phi \partial_d \phi - \frac{m^2}{2} n_I \phi^2 - \frac{n_I}{2 \det h} \Pi^2 \nonumber \\
&-& \frac{n_J \eta^{JK} \epsilon^{cd} \epsilon_{IKL} e_d^L}{\det h} \Pi \partial_c \phi \Big) \Big] \mathrm{d}^2 \sigma .
\end{eqnarray}

\subsection{Constraint analysis}

So let's start the constraint analysis. First, we must list all the constraints. The first constraints are the primary constraints. Explicitely, they read:
\begin{equation}
\left\{\begin{array}{rcl}
X^0_I &=& 0, \\
B^0_{IJ} &=& 0, \\
X^a_I &=& 0, \\
B^a_{IJ} &=& 2\alpha \epsilon_{IJK} \epsilon^{ab} e_b^K.
\end{array}\right.
\end{equation}
Their Poisson bracket with the Hamiltonian must be zero on shell so that the constraints are conserved. We will be using the following sign convention:
\begin{equation}
\{q, p\} = -1
\end{equation}
where $q$ represents a fundamental variable ($\phi$, $e$ or $A$) and $p$ the corresponding conjugated momentum ($\Pi$, $X$ or $B$). Let's study this. First:
\begin{equation}
\{X^0_I, H\} = -\alpha \epsilon_{IJK} \epsilon^{ab} F_{ab}^{JK}[A] - \Lambda n_I + \frac{1}{2} n_I h^{cd} \partial_c \phi \partial_d \phi + \frac{m^2}{2} n_I \phi^2 + \frac{n_I}{2 \det h} \Pi^2 + \frac{n_J \eta^{JK} \epsilon^{cd} \epsilon_{IKL} e_d^L}{\det h} \Pi \partial_c \phi .
\end{equation}
We will simply write this quantity $D_I$. No Lagrange multiplier appears here, so necessarily, $D_I = 0$, which is indeed the curvature constraint. Similarly:
\begin{equation}
\{B^0_{IJ}, H\} = 2\alpha \epsilon_{IJK} \epsilon^{ab} \partial_b e_a^I.
\end{equation}
Here, we can identify some version of the Gauß constraint, which we will write $G_{IJ} = 0$.

It is easy to see that no other constraint arise as the other commutators all involve Lagrange multipliers and can be inverted. Therefore the system of equations is now:
\begin{equation}
\left\{\begin{array}{rcl}
0 &=& X^0_I, \\
0 &=& B^0_{IJ}, \\
0 &=& X^a_I, \\
0 &=& B^a_{IJ} - 2\alpha \epsilon_{IJK} \epsilon^{ab} e_b^K, \\
0 &=& -\alpha \epsilon_{IJK} \epsilon^{ab} F_{ab}^{JK}[A] - \Lambda n_I + \frac{1}{2} n_I h^{cd} \partial_c \phi \partial_d \phi + \frac{m^2}{2} n_I \phi^2 + \frac{n_I}{2 \det h} \Pi^2 + \frac{n_J \eta^{JK} \epsilon^{cd} \epsilon_{IKL} e_d^L}{\det h} \Pi \partial_c \phi, \\
0 &=& 2 \alpha \epsilon_{IJK} \epsilon^{ab} \partial_b e_a^I.
\end{array}\right.
\end{equation}
The first two constraints are obviously first class. The last four are not, but that does not mean we have found all the first class constraints.

It can be checked that the following constraint is first class:
\begin{equation}
\partial_b B^b_{IJ} = 0 .
\end{equation}
It obvisouly commutes with every constraint and it is a constraint as a linear combination of the Gauß constraint found so far and the simplicity constraint. Finally, it is quite obvious that:
\begin{equation}
\alpha \epsilon_{IJK} \epsilon^{ab} F_{ab}^{JK}[A] + \Lambda \tilde{n}_I - \frac{1}{2} \tilde{n}_I \tilde{h}^{cd} \partial_c \phi \partial_d \phi - \frac{m^2}{2} \tilde{n}_I \phi^2 - \frac{\tilde{n}_I}{2 \det \tilde{h}} \Pi^2 - \frac{\tilde{n}_J \eta^{JK} \epsilon^{cd} \epsilon_{IKL} \tilde{e}_d^L}{\det \tilde{h}} \Pi \partial_c \phi = 0 
\end{equation}
where the tilded quantitites are constructed out of $B$ (rather than $e$), is a first class constraint.

Counting the number of degrees of freedom, we find that necessarily, the last constraints are second class. That is:
\begin{equation}
\left\{\begin{array}{rcl}
0 &=& X^a_I, \\
0 &=& B^a_{IJ} - 2\alpha \epsilon_{IJK} \epsilon^{ab} e_b^K, \\
\end{array}\right.
\end{equation}
are second class. This allows the computation of the Dirac brackets:
\begin{equation}
\left\{\begin{array}{rcl}
\{e^I_0(x), X_J^0(y)\}_D &=& -\delta^I_J \delta(x-y),\\
\{A^{IJ}_0(x), B_{KL}^0(y)\}_D &=& -(\delta^I_K \delta^J_L - \delta^I_L \delta^J_K) \delta(x-y),\\
\{A^{IJ}_a(x), e^{K}_b(y)\}_D &=& \frac{1}{2\alpha\det h} \epsilon_{ab} \epsilon^{IJK} \delta(x-y),\\
\{A^{IJ}_a(x), B_{KL}^b(y)\}_D &=& -\delta_a^b (\delta^I_K \delta^J_L - \delta^I_L \delta^J_K) \delta(x-y),\\
\{\phi(x), \Pi(y)\}_D &=& -\delta(x-y),
\end{array}\right.
\end{equation}
all other (non-fundamental) brackets being zero (including brackets dealing with $X_I^a$). With these brackets, it is rather obvious that the second class constraints commute with all the other constraints. Interestingly, they can be solved, and the system can finally be rewritten as:
\begin{equation}
\left\{
\begin{array}{rcl}
0 &=& \alpha \epsilon_{IJK} \epsilon^{ab} F_{ab}^{JK}[A] + \Lambda n_I - \frac{1}{2} n_I h^{cd} \partial_c \phi \partial_d \phi - \frac{m^2}{2} n_I \phi^2 - \frac{n_I}{2 \det h} \Pi^2 - \frac{n_J \eta^{JK} \epsilon^{cd} \epsilon_{IKL} e_d^L}{\det h} \Pi \partial_c \phi, \\
0 &=& \epsilon^{ab} \partial_b e_a^I,
\end{array}
\right.
\end{equation}
with the following brackets:
\begin{equation}
\left\{\begin{array}{rcl}
\{A^{IJ}_a(x), e^{K}_b(y)\} &=& \frac{1}{2\alpha\det h} \epsilon_{ab} \epsilon^{IJK} \delta(x-y),\\
\{\phi(x), \Pi(y)\} &=& -\delta(x-y).
\end{array}\right.
\end{equation}
The $B$ variables have been removed thanks to the second class constraints and the time component variables have been removed as they decouple from the rest and can be trivially solved. We now have the Hamiltonian formulation of our problem. This concludes this appendix.

\section{Brackets between the ladder operators and the constraints}
\label{app:operators}

In this appendix, we consider the bracket (using the Dirac bracket found in the previous appendix \ref{app:hamil}) between the curvature constraints and the would-be creation and annihilation operators. Namely, we want to compute $\{D_I, a_k\}_D$ (for which we will now drop the $D$ index from now on) where:
\begin{equation}
\left\{
\begin{array}{rcl}
D_I &=& \alpha \epsilon_{IJK} \epsilon^{ab} F_{ab}^{JK}[A] + \Lambda n_I - \frac{1}{2} n_I h^{cd} \partial_c \phi \partial_d \phi - \frac{m^2}{2} n_I \phi^2 - \frac{n_I}{2 \det h} \Pi^2 - \frac{n_J \eta^{JK} \epsilon^{cd} \epsilon_{IKL} e_d^L}{\det h} \Pi \partial_c \phi, \\
a_k &=& \frac{1}{\sqrt{2}\pi}\int \left(k^I n_I \phi + s \mathrm{i}\Pi\right) \mathrm{e}^{- \mathrm{i} \vec{k} \cdot \int^{\sigma} \vec{e}} \mathrm{d}^2\sigma ,
\end{array}
\right.
\end{equation}
where $s$ is a sign to be determined. To deal with this problem properly, we will need to integrate $D_I$ with a test field. We will therefore compute the following bracket:
\begin{equation}
\{\int N^I(\tau) D_I(\tau) \mathrm{d}^2\tau, a_k\}
\end{equation}
where both terms now have regular dependency on the variables and $N^I$ is the test field we just mentioned.

In this bracket, we can distinguish three kinds of terms, when expanding $D_I$. First, the bracket involving the cosmological constant term is trivial. Indeed, this terms only depends on the triad and as $a_k$ does not depend on $A$ at all, the bracket is zero. Second, we have the bracket involving the curvature of $A$. This part of $D_I$ does not depend on the matter field. As a consequence, only the dependence on the triad in $a_k$ will be of importance. Third, and finally, we will have the part of $D_I$ which involves the matter fields but does not involve the connection. And there only, the dependence on the matter fields in $a_k$ will be important for computing the brackets. The hope is of course that these last two terms compensate. It is quite intuitive that it is possible since this would correspond to $a_k$ creating energy on the matter field and compensating by giving the correct curvature to satisfy the Einstein equation.

Let's start by computing the following bracket:
\begin{equation}
A = \{\int N^I(\tau) \alpha \epsilon_{IJK} \epsilon^{ab} F_{ab}^{JK}[A](\tau) \mathrm{d}^2\tau, a_k\}.
\end{equation}
We have:
\begin{eqnarray}
A &=& \int N^I(\tau) \frac{\alpha}{2} \epsilon_{IJK} \epsilon^{ab} \{\partial_a A_b^{JK}(\tau) - \partial_b A_a^{JK}(\tau), a_k\} \mathrm{d}^2\tau \nonumber \\
&=& \int N^I(\tau) \alpha \epsilon_{IJK} \epsilon^{ab} \{\partial_a A_b^{JK}(\tau), a_k\} \mathrm{d}^2\tau \nonumber \\
&=& \int \int N^I(\tau) \frac{\alpha}{\sqrt{2}\pi} \epsilon_{IJK} \epsilon^{ab} \left(\phi(\sigma)\{\partial_a A_b^{JK}(\tau), k^L n_L(\sigma) \mathrm{e}^{- \mathrm{i} \vec{k} \cdot \int^{\sigma} \vec{e}}\} + s \mathrm{i}\Pi(\sigma)\{\partial_a A_b^{JK}(\tau), \mathrm{e}^{- \mathrm{i} \vec{k} \cdot \int^{\sigma} \vec{e}}\} \right)\mathrm{d}^2\sigma\mathrm{d}^2\tau \nonumber \\
&=& \int \int N^I(\tau) \frac{\alpha}{\sqrt{2}\pi} \epsilon_{IJK} \epsilon^{ab} \left(\phi(\sigma)k^L \{\partial_a A_b^{JK}(\tau), n_L(\sigma)\} \mathrm{e}^{- \mathrm{i} \vec{k} \cdot \int^{\sigma} \vec{e}} + (k^L n_L(\sigma) \phi(\sigma) + s\mathrm{i}\Pi(\sigma))\{\partial_a A_b^{JK}(\tau), \mathrm{e}^{- \mathrm{i} \vec{k} \cdot \int^{\sigma} \vec{e}}\} \right)\mathrm{d}^2\sigma\mathrm{d}^2\tau \nonumber \\
& & 
\end{eqnarray}
Let's compute the two intermediary brackets. First, we have:
\begin{eqnarray}
\{\partial_a A_b^{JK}(\tau), n_L(\sigma)\} &=& \frac{\partial}{\partial \tau^a}\{ A_b^{JK}(\tau), \frac{1}{2} \epsilon_{LMN} \epsilon^{cd} e_c^M(\sigma) e_d^N(\sigma)\} \nonumber \\
&=& \epsilon_{LMN} \epsilon^{cd} e_c^M(\sigma) \frac{\partial}{\partial \tau^a}\{ A_b^{JK}(\tau), e_d^N(\sigma)\} \nonumber \\
&=& \epsilon_{LMN} \epsilon^{cd} e_c^M(\sigma) \frac{\partial}{\partial \tau^a} \left( \frac{1}{2\alpha\det h(\sigma)} \epsilon_{bd}(\sigma) \epsilon^{JKN} \delta(\tau-\sigma)  \right) \nonumber \\
&=& \frac{1}{2\alpha\det h(\sigma)} \epsilon_{bd}(\sigma) \epsilon^{NJK} \epsilon_{NLM} \epsilon^{cd} e_c^M(\sigma) \frac{\partial}{\partial \tau^a} \left(  \delta(\tau-\sigma) \right) \nonumber \\
&=& \frac{1}{2\alpha\det h(\sigma)} h_{bb'}(\sigma) \epsilon^{cd} h_{dd'}(\sigma) \epsilon^{b'd'} (\delta^J_M \delta^K_L - \delta^J_L \delta^K_M)  e_c^M(\sigma) \frac{\partial}{\partial \tau^a} \left(  \delta(\tau-\sigma) \right) \nonumber \\
&=& \frac{1}{2\alpha\det h(\sigma)} h_{bb'}(\sigma) (\det h(\sigma)) h^{cb'}(\sigma) (\delta^J_M \delta^K_L - \delta^J_L \delta^K_M) e_c^M(\sigma) \frac{\partial}{\partial \tau^a} \left(  \delta(\tau-\sigma) \right) \nonumber \\
&=& \frac{1}{2\alpha} (\delta^J_M \delta^K_L - \delta^J_L \delta^K_M) e_b^M(\sigma) \frac{\partial}{\partial \tau^a} \left(  \delta(\tau-\sigma) \right) \nonumber \\
&=& \frac{1}{2\alpha} (\delta^J_L \delta^K_M - \delta^J_M \delta^K_L) e_b^M(\sigma) \frac{\partial}{\partial \sigma^a} \left(  \delta(\tau-\sigma) \right) .
\end{eqnarray}
We used the equality between the two derivatives for $\delta$ (up to a sign) on the last line to avoid the appearance of derivatives of $N^I$ in the full bracket.

Now, we also have (we include the initial $\epsilon$ for simplifications):
\begin{eqnarray}
\epsilon_{IJK} \epsilon^{ab} \{\partial_a A_b^{JK}(\tau), \mathrm{e}^{- \mathrm{i} \vec{k} \cdot \int^{\sigma} \vec{e}}\} &=& \epsilon_{IJK} \epsilon^{ab}\frac{\partial}{\partial \tau^a} \{ A_b^{JK}(\tau), \mathrm{e}^{- \mathrm{i} \vec{k} \cdot \int^{\sigma} \vec{e}}\} \nonumber \\
&=& \epsilon_{IJK} \epsilon^{ab}\int_\xi \left( - \int^\sigma \mathrm{i} k^P \eta_{PQ} \delta(\xi - \zeta(s) )\frac{\mathrm{d}\zeta^c}{\mathrm{d}s} \mathrm{d}s \right) \mathrm{e}^{- \mathrm{i} \vec{k} \cdot \int^{\sigma} \vec{e}} \frac{\partial}{\partial \tau^a} \{A_b^{JK}(\tau), e_c^Q(\xi) \} \mathrm{d}^2\xi \nonumber \\
&=& \epsilon_{IJK} \epsilon^{ab}\int_\xi \left( - \int^\sigma \mathrm{i} k^P \eta_{PQ} \delta(\xi - \zeta(s) )\frac{\mathrm{d}\zeta^c}{\mathrm{d}s} \mathrm{d}s \right) \mathrm{e}^{- \mathrm{i} \vec{k} \cdot \int^{\sigma} \vec{e}} \nonumber \\
&\times& \frac{\partial}{\partial \tau^a} \left( \frac{1}{2\alpha \det h(\xi)}\epsilon_{bc}(\xi) \epsilon^{JKQ} \delta(\tau - \xi) \right)  \mathrm{d}^2\xi \nonumber \\
&=& -\frac{1}{\alpha} \int_\xi \left( \int^\sigma \mathrm{i} k^P \eta_{PI} \delta(\xi - \zeta(s) )\frac{\mathrm{d}\zeta^a}{\mathrm{d}s} \mathrm{d}s \right) \mathrm{e}^{- \mathrm{i} \vec{k} \cdot \int^{\sigma} \vec{e}} \frac{\partial}{\partial \tau^a} \left( \delta(\tau - \xi) \right)  \mathrm{d}^2\xi \nonumber \\
&=& \frac{1}{\alpha} \int_\xi \left( \int^\sigma \mathrm{i} k^P \eta_{PI} \delta(\xi - \zeta(s) )\frac{\mathrm{d}\zeta^a}{\mathrm{d}s} \mathrm{d}s \right) \mathrm{e}^{- \mathrm{i} \vec{k} \cdot \int^{\sigma} \vec{e}} \frac{\partial}{\partial \xi^a} \left( \delta(\tau - \xi) \right)  \mathrm{d}^2\xi \nonumber \\
&=& -\frac{1}{\alpha} \int_\xi \left( \int^\sigma \mathrm{i} k^P \eta_{PI} \frac{\partial}{\partial \xi^a} (\delta(\xi - \zeta(s) ))\frac{\mathrm{d}\zeta^a}{\mathrm{d}s} \mathrm{d}s \right) \mathrm{e}^{- \mathrm{i} \vec{k} \cdot \int^{\sigma} \vec{e}}  \delta(\tau - \xi)  \mathrm{d}^2\xi \nonumber \\
&=& \frac{1}{\alpha} \int_\xi \left( \int^\sigma \mathrm{i} k^P \eta_{PI} \frac{\mathrm{d}}{\mathrm{d} s} (\delta(\xi - \zeta(s) )) \mathrm{d}s \right) \mathrm{e}^{- \mathrm{i} \vec{k} \cdot \int^{\sigma} \vec{e}}  \delta(\tau - \xi)  \mathrm{d}^2\xi \nonumber \\
&=& \frac{\mathrm{i} k^P \eta_{PI}}{\alpha} \left( \int^\sigma  \frac{\mathrm{d}}{\mathrm{d} s} (\delta(\tau - \zeta(s) )) \mathrm{d}s \right) \mathrm{e}^{- \mathrm{i} \vec{k} \cdot \int^{\sigma} \vec{e}} \nonumber \\
&=& \frac{\mathrm{i} k^P \eta_{PI}}{\alpha} \delta(\tau - \sigma ) \mathrm{e}^{- \mathrm{i} \vec{k} \cdot \int^{\sigma} \vec{e}}
\end{eqnarray}
The last line should also contain an opposite contribution from the start point of the integral. To make this omission rigorous, we have to consider that $N^I$ has compact support. In that case, once the start point is sufficiently far, its contribution will always be zero. This however means that we have some restrictions on the distribution spaces we might consider.

Let's put all these computations together. We get:
\begin{eqnarray}
A &=& \int \int N^I(\tau) \frac{\alpha}{\sqrt{2}\pi} \epsilon_{IJK} \epsilon^{ab} \left(\phi(\sigma)k^L \{\partial_a A_b^{JK}(\tau), n_L(\sigma)\} \mathrm{e}^{- \mathrm{i} \vec{k} \cdot \int^{\sigma} \vec{e}} + (k^L n_L(\sigma) \phi(\sigma) + s\mathrm{i}\Pi(\sigma))\{\partial_a A_b^{JK}(\tau), \mathrm{e}^{- \mathrm{i} \vec{k} \cdot \int^{\sigma} \vec{e}}\} \right)\mathrm{d}^2\sigma\mathrm{d}^2\tau \nonumber \\
&=& \int \int N^I(\tau) \frac{\alpha}{\sqrt{2}\pi} \left(\epsilon_{IJK} \epsilon^{ab} \phi(\sigma)k^L \frac{1}{2\alpha} (\delta^J_L \delta^K_M - \delta^J_M \delta^K_L) e_b^M(\sigma) \frac{\partial}{\partial \sigma^a} \left(  \delta(\tau-\sigma) \right) \mathrm{e}^{- \mathrm{i} \vec{k} \cdot \int^{\sigma} \vec{e}} \right. \nonumber \\
&+& \left. (k^L n_L(\sigma) \phi(\sigma) + s\mathrm{i}\Pi(\sigma)) \frac{\mathrm{i} k^P \eta_{PI}}{\alpha} \delta(\tau - \sigma ) \mathrm{e}^{\mathrm{i} \vec{k} \cdot \int^{\sigma} \vec{e}} \right)\mathrm{d}^2\sigma\mathrm{d}^2\tau \nonumber \\
&=& \int \int \frac{N^I(\tau)}{\sqrt{2}\pi} \left(\epsilon_{IJK} \epsilon^{ab} \phi(\sigma)k^J e_b^K(\sigma) \frac{\partial}{\partial \sigma^a} \left(  \delta(\tau-\sigma) \right) \mathrm{e}^{- \mathrm{i} \vec{k} \cdot \int^{\sigma} \vec{e}} \right) \mathrm{d}^2\sigma\mathrm{d}^2\tau \nonumber \\
&+& \int \frac{ N^I}{\sqrt{2}\pi} \left( (k^L n_L \phi + s\mathrm{i}\Pi) \mathrm{i} k^P \eta_{PI} \mathrm{e}^{- \mathrm{i} \vec{k} \cdot \int^{\sigma} \vec{e}} \right)\mathrm{d}^2\sigma \nonumber \\
&\approx& \int \frac{N^I}{\sqrt{2}\pi} \left(\epsilon_{IJK} \epsilon^{ab} (-\partial_a \phi) k^J e_b^K + \epsilon_{IJK} \epsilon^{ab} \phi k^J e_b^K (\mathrm{i} k^P \eta_{PQ} e^Q_a) + (k^L n_L \phi + s\mathrm{i}\Pi) \mathrm{i} k^P \eta_{PI}\right) \mathrm{e}^{- \mathrm{i} \vec{k} \cdot \int^{\sigma} \vec{e}} \mathrm{d}^2\sigma \nonumber \\
&\approx& \int \frac{N^I}{\sqrt{2}\pi} \left(-(\epsilon_{IJK} \epsilon^{ab} k^J e_b^K)\partial_a \phi + (\epsilon_{IJK} \epsilon^{ab} \eta_{PQ} k^J k^P e_b^K e^Q_a + \eta_{PI}  k^L k^P n_L ) \mathrm{i}\phi - s(\eta_{PI} k^P) \Pi \right) \mathrm{e}^{- \mathrm{i} \vec{k} \cdot \int^{\sigma} \vec{e}} \mathrm{d}^2\sigma ,
\end{eqnarray}
where everything with the $\approx$ is only true on-shell and more precisely when the Gauß constraints are verified.

Let's now turn to the second half of the computation:
\begin{equation}
\begin{array}{c}
B= \\
\{ \int N^I(\tau) \left(- \frac{1}{2} n_I(\tau) h^{cd}(\tau) \partial_c \phi(\tau) \partial_d \phi(\tau) - \frac{m^2}{2} n_I(\tau) \phi(\tau)^2 + \frac{n_I(\tau)}{2 \det h(\tau)} \Pi(\tau)^2 + \frac{n_J(\tau) \eta^{JK} \epsilon^{cd} \epsilon_{IKL} e_d^L(\tau)}{\det h(\tau)} \Pi(\tau) \partial_c \phi(\tau) \right) \mathrm{d}^2\tau , a_k\} .
\end{array}
\end{equation}
Once more, let's split this expression into simpler components. We will have:
\begin{eqnarray}
B_1 &=& \{ \int N^I(\tau) \left(- \frac{1}{2} n_I(\tau) h^{cd}(\tau) \partial_c \phi(\tau) \partial_d \phi(\tau) \right) \mathrm{d}^2\tau , a_k\}, \\
B_2 &=& \{ \int N^I(\tau) \left(- \frac{m^2}{2} n_I(\tau) \phi(\tau)^2 \right) \mathrm{d}^2\tau , a_k\}, \\
B_3 &=& \{ \int N^I(\tau) \left(-\frac{n_I(\tau)}{2 \det h(\tau)} \Pi(\tau)^2 \right) \mathrm{d}^2\tau , a_k\}, \\
B_4 &=& \{ \int N^I(\tau) \left(-\frac{n_J(\tau) \eta^{JK} \epsilon^{cd} \epsilon_{IKL} e_d^L(\tau)}{\det h(\tau)} \Pi(\tau) \partial_c \phi(\tau) \right) \mathrm{d}^2\tau , a_k\}.
\end{eqnarray}
Let's compute each one of them separately, starting with $B_1$:
\begin{eqnarray}
B_1 &=& \{ \int N^I(\tau) \left(- \frac{1}{2} n_I(\tau) h^{cd}(\tau) \partial_c \phi(\tau) \partial_d \phi(\tau) \right) \mathrm{d}^2\tau , a_k\} \nonumber \\
&=& - \int N^I(\tau) n_I(\tau) h^{cd}(\tau) \partial_c \phi(\tau) \{ \partial_d \phi(\tau) , a_k\}  \mathrm{d}^2\tau \nonumber \\
&=& - \int N^I(\tau) n_I(\tau) h^{cd}(\tau) \partial_c \phi(\tau) \frac{\partial}{\partial \tau^d} \{ \phi(\tau) , a_k\} \mathrm{d}^2\tau
\end{eqnarray}
This calls for the following computation:
\begin{eqnarray}
\{ \phi(\tau) , a_k\} &=& \{ \phi(\tau) , \frac{1}{\sqrt{2}\pi}\int \left(k^I n_I(\sigma) \phi(\sigma) + s\mathrm{i}\Pi(\sigma)\right) \mathrm{e}^{- \mathrm{i} \vec{k} \cdot \int^{\sigma} \vec{e}} \mathrm{d}^2\sigma \}  \nonumber \\
&=& \frac{\mathrm{i}s}{\sqrt{2}\pi}\int  \{ \phi(\tau), \Pi(\sigma)\} \mathrm{e}^{- \mathrm{i} \vec{k} \cdot \int^{\sigma} \vec{e}} \mathrm{d}^2\sigma \nonumber \\
&=& - \frac{\mathrm{i}s}{\sqrt{2}\pi}\int \delta(\tau - \sigma) \mathrm{e}^{- \mathrm{i} \vec{k} \cdot \int^{\sigma} \vec{e}} \mathrm{d}^2\sigma \nonumber \\
&=& - \frac{\mathrm{i}s}{\sqrt{2}\pi} \mathrm{e}^{- \mathrm{i} \vec{k} \cdot \int^{\tau} \vec{e}}
\end{eqnarray}
Putting it back into $B_1$, we get:
\begin{eqnarray}
B_1 &=& - \int N^I(\tau) n_I(\tau) h^{cd}(\tau) \partial_c \phi(\tau) \frac{\partial}{\partial \tau^d} \{ \phi(\tau) , a_k\} \mathrm{d}^2\tau \nonumber \\
&=& - \int N^I(\tau) n_I(\tau) h^{cd}(\tau) \partial_c \phi(\tau) \frac{\partial}{\partial \tau^d} \left( - \frac{\mathrm{i}s}{\sqrt{2}\pi} \mathrm{e}^{- \mathrm{i} \vec{k} \cdot \int^{\tau} \vec{e}} \right) \mathrm{d}^2\tau \nonumber \\
&=& \frac{s}{\sqrt{2}\pi} \int N^I n_I h^{cd} \partial_c \phi k^P \eta_{PQ} e^Q_d \mathrm{e}^{- \mathrm{i} \vec{k} \cdot \int^{\sigma} \vec{e}} \mathrm{d}^2 \sigma \nonumber \\
&=& \int \frac{N^I}{\sqrt{2}\pi} (s \eta_{PQ} k^P h^{ab} n_I e^Q_b) \partial_a \phi \mathrm{e}^{- \mathrm{i} \vec{k} \cdot \int^{\sigma} \vec{e}} \mathrm{d}^2 \sigma ,
\end{eqnarray}
where the last expression was written in a form similar to that of $A$.

Let's now consider $B_2$:
\begin{eqnarray}
B_2 &=& \{ \int N^I(\tau) \left(- \frac{m^2}{2} n_I(\tau) \phi(\tau)^2 \right) \mathrm{d}^2\tau , a_k\} \nonumber \\
&=& - \int N^I(\tau) m^2 n_I(\tau) \phi(\tau) \{\phi(\tau) , a_k\} \mathrm{d}^2 \tau \nonumber \\
&=& - \int N^I(\tau) m^2 n_I(\tau) \phi(\tau) \left( - s\frac{\mathrm{i}}{\sqrt{2}\pi} \mathrm{e}^{- \mathrm{i} \vec{k} \cdot \int^{\tau} \vec{e}} \right) \mathrm{d}^2 \tau \nonumber \\
&=& \int \frac{N^I}{\sqrt{2}\pi} (s m^2 n_I) \mathrm{i}\phi \mathrm{e}^{- \mathrm{i} \vec{k} \cdot \int^{\sigma} \vec{e}} \mathrm{d}^2 \sigma .
\end{eqnarray}

Let's turn to $B_3$:
\begin{eqnarray}
B_3 &=& \{ \int N^I(\tau) \left(-\frac{n_I(\tau)}{2 \det h(\tau)} \Pi(\tau)^2 \right) \mathrm{d}^2\tau , a_k\} \nonumber \\
&=& -\int N^I(\tau) \frac{n_I(\tau)}{\det h(\tau)} \Pi(\tau) \{\Pi(\tau), a_k \} \mathrm{d}^2\tau .
\end{eqnarray}
We must now compute:
\begin{eqnarray}
\{ \Pi(\tau) , a_k\} &=& \{ \Pi(\tau) , \frac{1}{\sqrt{2}\pi}\int \left(k^I n_I(\sigma) \phi(\sigma) + s\mathrm{i}\Pi(\sigma)\right) \mathrm{e}^{- \mathrm{i} \vec{k} \cdot \int^{\sigma} \vec{e}} \mathrm{d}^2\sigma \}  \nonumber \\
&=& \frac{1}{\sqrt{2}\pi}\int k^I n_I(\sigma) \{ \Pi(\tau), \phi(\sigma)\} \mathrm{e}^{- \mathrm{i} \vec{k} \cdot \int^{\sigma} \vec{e}} \mathrm{d}^2\sigma \nonumber \\
&=& \frac{1}{\sqrt{2}\pi}\int k^I n_I(\sigma) \delta(\tau - \sigma) \mathrm{e}^{- \mathrm{i} \vec{k} \cdot \int^{\sigma} \vec{e}} \mathrm{d}^2\sigma \nonumber \\
&=& \frac{k^L n_L(\tau)}{\sqrt{2}\pi} \mathrm{e}^{- \mathrm{i} \vec{k} \cdot \int^{\tau} \vec{e}} .
\end{eqnarray}
This gives:
\begin{eqnarray}
B_3 &=& -\int N^I(\tau) \frac{n_I(\tau)}{\det h(\tau)} \Pi(\tau) \{\Pi(\tau), a_k \} \mathrm{d}^2\tau \nonumber \\
&=& -\int N^I(\tau) \frac{n_I(\tau)}{\det h(\tau)} \Pi(\tau) \left( \frac{k^L n_L(\tau)}{\sqrt{2}\pi} \mathrm{e}^{- \mathrm{i} \vec{k} \cdot \int^{\tau} \vec{e}} \right) \mathrm{d}^2\tau \nonumber \\
&=& \int \frac{N^I}{\sqrt{2}\pi} \left(-\frac{k^L n_L n_I}{\det h}\right) \Pi \mathrm{e}^{- \mathrm{i} \vec{k} \cdot \int^{\sigma} \vec{e}} \mathrm{d}^2 \sigma .
\end{eqnarray}

Finally, let's turn to $B_4$:
\begin{eqnarray}
B_4 &=& \{ \int N^I(\tau) \left(-\frac{n_J(\tau) \eta^{JK} \epsilon^{cd} \epsilon_{IKL} e_d^L(\tau)}{\det h(\tau)} \Pi(\tau) \partial_c \phi(\tau) \right) \mathrm{d}^2\tau , a_k\} \nonumber \\
&=& -\int N^I(\tau) \frac{n_J(\tau) \eta^{JK} \epsilon^{cd} \epsilon_{IKL} e_d^L(\tau)}{\det h(\tau)} \left( \{\Pi(\tau), a_k \}\partial_c \phi(\tau) + \Pi(\tau) \frac{\partial}{\partial \tau^c}\{\phi(\tau), a_k \}\right)  \mathrm{d}^2\tau \nonumber \\
&=& -\int N^I(\tau) \frac{n_J(\tau) \eta^{JK} \epsilon^{cd} \epsilon_{IKL} e_d^L(\tau)}{\det h(\tau)} \left( \left( \frac{k^M n_M(\tau)}{\sqrt{2}\pi} \mathrm{e}^{- \mathrm{i} \vec{k} \cdot \int^{\tau} \vec{e}} \right)\partial_c \phi(\tau) + \Pi(\tau) \frac{\partial}{\partial \tau^c}\left( - \frac{s\mathrm{i}}{\sqrt{2}\pi} \mathrm{e}^{- \mathrm{i} \vec{k} \cdot \int^{\tau} \vec{e}} \right)\right) \mathrm{d}^2\tau \nonumber \\
&=& \int \frac{N^I}{\sqrt{2}\pi} \left(\left[-\frac{n_J \eta^{JK} \epsilon^{ad} \epsilon_{IKL} e_d^L}{\det h} k^M n_M \right]\partial_a \phi + \left[s\frac{n_J \eta^{JK} \epsilon^{cd} \epsilon_{IKL} e_d^L}{\det h} k^P \eta_{PQ} e^Q_c \right]\Pi\right) \mathrm{e}^{- \mathrm{i} \vec{k} \cdot \int^{\sigma} \vec{e}}\mathrm{d}^2\sigma
\end{eqnarray}

Before moving to the full expression, let's try and simplify the terms in $\partial_a \phi$ on one side and $\Pi$ on the other. First, for $\partial_a \phi$, we have:
\begin{equation}
C_1 = s\eta_{PQ} k^P h^{ab} n_I e^Q_b - \frac{n_J \eta^{JK} \epsilon^{ad} \epsilon_{IKL} e_d^L}{\det h} k^M n_M .
\end{equation}
And for, $\Pi$, we have:
\begin{equation}
C_2 = \frac{-k^L n_L n_I}{\det h} + s\frac{n_J \eta^{JK} \epsilon^{cd} \epsilon_{IKL} e_d^L}{\det h} k^P \eta_{PQ} e^Q_c .
\end{equation}

$C_2$ is slightly simpler, let's start with it. Indeed, we now we'd like to find $-s\eta_{PI} k^P$ so that it exactly compensates the term in $A$. So let's compute:
\begin{equation}
\eta_{PI} k^P = \delta^J_I \eta_{PJ} k^P .
\end{equation}
We will now try to find another way to write $\delta^J_I$. For this, let's consider the tetrad $d$ defined by, for all spatial directions $a$, $d_a^I = e_a^I$ and for the time direction, $d_0^I = \frac{\eta^{IJ} n_J}{\sqrt{-n^2}}$ (where $n^2$ is the Minkowski square of $n_I$). If the triad is non-degenerate (which we assumed), $d$ is invertible by construction and $\det d = -\sqrt{-n^2}$. Therefore, we can write:
\begin{eqnarray}
\delta^J_I &=& d^J_\mu d^\mu_I \nonumber \\
&=& d^J_0 d^0_I + d^J_a d^a_I \nonumber \\
&=& \frac{\eta^{JR} n_R}{\sqrt{-n^2}} \frac{\epsilon_{IMN} \epsilon^{cd} e_c^M e_d^N}{2 (-\sqrt{-n^2})} + e_a^J \frac{\epsilon^{ab} \epsilon_{IMN} d_b^M d_0^N}{(-\sqrt{-n^2})} \nonumber \\
&=& \frac{\eta^{JR} n_R n_I}{n^2} + \frac{\epsilon^{ab} \epsilon_{IMN} e_a^J e_b^M \eta^{NL} n_L}{n^2} \nonumber \\
&=& - \frac{\eta^{JR} n_R n_I}{\det h} - \frac{\epsilon^{ab} \epsilon_{IMN} e_a^J e_b^M \eta^{NL} n_L}{\det h} .
\end{eqnarray}
The last line uses $\det h = -n^2$. Therefore:
\begin{eqnarray}
\eta_{PI} k^P &=& \eta_{PJ} \delta^J_I k^P \nonumber \\
&=& -\eta_{PJ} k^P \left(\frac{\eta^{JR} n_R n_I}{\det h} + \frac{\epsilon^{ab} \epsilon_{IMN} e_a^J e_b^M \eta^{NL} n_L}{\det h}\right) \nonumber \\
&=& -\frac{k^L n_L n_I}{\det h} - \frac{\epsilon_{ILK} e^L_d \eta^{KJ} n_J \epsilon^{cd} e_c^Q}{\det h} \eta_{PQ} k^P \nonumber \\
&=& -\frac{k^L n_L n_I}{\det h} + \frac{\epsilon_{IKL} e^L_d \eta^{JK} n_J \epsilon^{cd} e_c^Q}{\det h} \eta_{PQ} k^P
\end{eqnarray}
And so, we get (for $s=1$):
\begin{equation}
C_2 = s\eta_{PI} k^P ,
\end{equation}
which is exactly what we wanted.

Let's turn to $C_1$. Once more, we know what we would like. We would like to compensate the term $-\epsilon_{IJK}\epsilon^{ab}k^J e_b^K$ coming from $A$. So, we would like $C_1$ to be equal to the opposite. Once more, let's start from the desired expression:
\begin{eqnarray}
\epsilon_{IJK}\epsilon^{ab}k^J e_b^K &=& \epsilon_{IJK}\epsilon^{ab} \delta^J_S k^S e_b^K \nonumber \\
&=& -\epsilon_{IJK}\epsilon^{ab} \left(\frac{\eta^{JR} n_R n_S}{\det h} + \frac{\epsilon^{cd} \epsilon_{SMN} e_c^J e_d^M \eta^{NL} n_L}{\det h}\right) k^S e_b^K \nonumber \\
&=& -\epsilon_{IJK}\epsilon^{ab} \frac{\eta^{JR} n_R n_S}{\det h}  k^S e_b^K - \epsilon_{IJK}\epsilon^{ab} \frac{\epsilon^{cd} \epsilon_{SMN} e_c^J e_d^M \eta^{NL} n_L}{\det h}  k^S e_b^K \nonumber \\
&=& -\frac{n_J \eta^{JK}\epsilon^{ad}\epsilon_{IKL} e_d^L}{\det h} k^M n_M  - \epsilon_{IJK}\epsilon^{ab} \frac{\epsilon^{cd} \epsilon_{SMN} e_c^J e_d^M \eta^{NL} n_L}{\det h}  k^S e_b^K \nonumber \\
&=& -\frac{n_J \eta^{JK}\epsilon^{ad}\epsilon_{IKL} e_d^L}{\det h} k^M n_M  - \epsilon_{IJK}\epsilon^{ab} \frac{\epsilon^{cd} \epsilon_{SMN} e_c^J e_d^M \eta^{NL} \epsilon_{LPQ} \epsilon^{ij} e_i^P e_j^Q}{2\det h}  k^S e_b^K \nonumber \\
&=& -\frac{n_J \eta^{JK}\epsilon^{ad}\epsilon_{IKL} e_d^L}{\det h} k^M n_M  - \epsilon_{IJK}\epsilon^{ab} \frac{\epsilon^{cd} (\eta_{SQ} \eta_{MP} - \eta_{SP} \eta_{MQ}) e_c^J e_d^M \epsilon^{ij} e_i^P e_j^Q}{2\det h}  k^S e_b^K \nonumber \\
&=& -\frac{n_J \eta^{JK}\epsilon^{ad}\epsilon_{IKL} e_d^L}{\det h} k^M n_M  - \epsilon_{IJK}\epsilon^{ab} \frac{\epsilon^{cd} \eta_{SQ} e_c^J \epsilon^{ij} h_{id} e_j^Q}{\det h}  k^S e_b^K \nonumber \\
&=& -\frac{n_J \eta^{JK}\epsilon^{ad}\epsilon_{IKL} e_d^L}{\det h} k^M n_M  + \epsilon_{IJK}\epsilon^{ab} \eta_{SQ} e_c^J h^{jc} e_j^Q  k^S e_b^K \nonumber \\
&=& -\frac{n_J \eta^{JK}\epsilon^{ad}\epsilon_{IKL} e_d^L}{\det h} k^M n_M  + \epsilon_{IJK}e_{c'}^J e_{b'}^K \frac{\epsilon^{b'c'} \epsilon_{bc}}{2\det h} \epsilon^{ab} \eta_{SQ} h^{jc} e_j^Q  k^S  \nonumber \\
&=& -\frac{n_J \eta^{JK}\epsilon^{ad}\epsilon_{IKL} e_d^L}{\det h} k^M n_M  - \frac{\epsilon_{dc}}{\det h} n_I\epsilon^{ad} \eta_{PQ} h^{bc} e_b^Q  k^P  \nonumber \\
&=& -\frac{n_J \eta^{JK}\epsilon^{ad}\epsilon_{IKL} e_d^L}{\det h} k^M n_M  +  n_I \eta_{PQ} h^{ba} e_b^Q  k^P
\end{eqnarray}
And so, we get (once more for $s=1$):
\begin{equation}
C_1 = \epsilon_{IJK}\epsilon^{ab}k^J e_b^K
\end{equation}
which is once again what we wanted.

The only thing that remains is the term in $\phi$. This time it is more natural to look at the term in $A$ and try to get the necessary term to compensate in $B$, namely to compensate $sm^2n_I$. We have:
\begin{eqnarray}
D &=& \epsilon_{IJK} \epsilon^{ab} \eta_{PQ} k^J k^P e_b^K e^Q_a + \eta_{PI}  k^L k^P n_L \nonumber \\
&=& \epsilon_{IJK} \eta_{PQ} k^J k^P \epsilon^{ab} e_b^{K'} e^{Q'}_a \frac{\delta_{K'}^K \delta_{Q'}^Q -\delta_{K'}^Q \delta_{Q'}^K}{2} + \eta_{PI}  k^L k^P n_L \nonumber \\
&=& \epsilon_{IJK} \eta_{PQ} k^J k^P \epsilon^{ab} e_b^{K'} e^{Q'}_a \frac{\epsilon^{LKQ} \epsilon_{LQ'K'}}{2} + \eta_{PI}  k^L k^P n_L \nonumber \\
&=& \epsilon_{IJK} \eta_{PQ} k^J k^P n_L \epsilon^{LKQ} + \eta_{PI}  k^L k^P n_L \nonumber \\
&=& (\delta^L_I \delta^Q_J - \delta^L_J \delta^Q_I) \eta_{PQ} k^J k^P n_L + \eta_{PI}  k^L k^P n_L \nonumber \\
&=& \eta_{PJ} k^J k^P n_I - \eta_{PI} k^J k^P n_J + \eta_{PI}  k^L k^P n_L \nonumber \\
&=& k^2 n_I \nonumber \\
&=& -m^2 n_I
\end{eqnarray}
which is indeed $-sm^2 n_I$ for $s=1$. Putting all this together, we do get:
\begin{equation}
A+B_1 + B_2 + B_3 + B_4 \approx 0.
\end{equation}
Or to put it in the original question terms:
\begin{equation}
\{\int N^I(\tau) D_I(\tau) \mathrm{d}^2\tau, a_k\} \approx 0
\end{equation}
if $s=1$. It is to be noted that this result holds on-shell, when the Gauß constraint is checked. Otherwise, the bracket is linear in the Gauß constraints.

\bibliographystyle{bib-style}
\bibliography{biblio}

\end{document}